\documentclass[web]{ieeecolor}
\newcommand{\logoname}{}
\usepackage[font=small,labelfont=bf]{caption}
\usepackage{generic}
\usepackage{cite}
\usepackage{amsmath,amssymb,amsfonts}
\usepackage{algorithmic}
\usepackage{graphicx}
\usepackage{algorithm,algorithmic}
\usepackage{hyperref}
\hypersetup{hidelinks}
\usepackage{textcomp}
\usepackage{adjustbox}
\usepackage[table,rgb,dvipsnames]{xcolor}   
\usepackage{booktabs}
\usepackage{multirow}
\usepackage{subcaption}
\definecolor{sigcol}{rgb}{0.8, 1.0, 0.8} 
\begin{document}
\title{Load-Aware Calibration of EMG-Driven Musculoskeletal Models for Accurate and Generalizable Joint Torque Estimation}
\author{Rajnish Kumar, Suriya Prakash Muthukrishnan, Lalan Kumar, and Sitikantha Roy
\thanks{Rajnish Kumar is with Department of Applied Mechanics, Indian Institute of Technology (IIT) Delhi, New Delhi 110016 India (e-mail: amz208469@iitd.ac.in).}
\thanks{Suriya Prakash Muthukrishnan is with Department of Physiology at All India Institute of Medical Sciences (AIIMS), New Delhi 110608 India (e-mail: dr.suriyaprakash@aiims.edu).}
\thanks{Lalan Kumar is with Department of Electrical Engineering, IIT Delhi, New Delhi 110016 India (e-mail: lalank@ee.iitd.ac.in).}
\thanks{S. Roy is with Department of Applied Mechanics, IIT Delhi, New Delhi 110016 India (e-mail: sroy@am.iitd.ac.in).}}

\maketitle

\begin{abstract}
Accurate EMG-driven musculoskeletal (MSK) modeling is critical for biomechanics, rehabilitation, and assistive technology. However, most models calibrate parameters under a single load, ignoring the fact that tasks with similar kinematics may differ in mechanical demand. This study introduces a load-aware calibration framework to improve joint torque prediction accuracy and generalizability.
Surface EMG and joint kinematics were recorded from eleven participants during elbow flexion-extension under 0, 2, and 4\,kg loads. We evaluated three calibration strategies (load-specific, global, cross-load) and three optimization frameworks (simulated annealing (SA), particle swarm optimization (PSO), and hybrid PSO-pattern search (PSO-PS)).
Results indicate that load-specific calibration significantly improves performance, with lower RMSE and higher correlation ($r > 0.75$). Parameters related to muscle force, fiber length, and activation dynamics showed high load sensitivity. PSO-based methods yielded more consistent and physiologically plausible estimates than simulated annealing.
The proposed framework enables MSK models to distinguish between visually similar but mechanically distinct tasks, supporting robust subject-specific modeling for clinical and real-world applications.
\end{abstract}

\section{Introduction}
\label{sec:introduction}
\IEEEPARstart{T}{he} computational modeling of human movement using musculoskeletal (MSK) frameworks has become foundational to biomechanics, motor neuroscience, and rehabilitation engineering~\cite{Buchanan2004neuromusculoskeletal,lloyd2023history}. 
By integrating anatomical structure, neural activation, and joint kinematics, these models enable the estimation of internal biomechanical variables, such as muscle forces and joint torques, that are otherwise inaccessible via non-invasive methods. 
Accurate MSK modeling is essential for scientific insight, personalized clinical decision-making, rehabilitation design, and the development of robotic or assistive technologies~\cite{Buchanan2004neuromusculoskeletal,crouch2016lumped,van2022building}. 
This has led to growing demand for robust, subject-specific modeling approaches that generalize well across contexts.

A key advancement in this field has been the transition from generic, scaled MSK models to personalized, EMG-informed frameworks. 
Traditional techniques often relied on static optimization and generic anatomical scaling to resolve muscle redundancy. However, such methods struggle to capture the substantial inter-individual variability in neuromuscular control~\cite{koo2002vivo,lloyd2023history}. 
EMG-driven MSK models address this gap by directly incorporating subject-specific neural inputs, enabling simulations that more accurately reflect physiological behavior~\cite{Buchanan2004neuromusculoskeletal,lloyd2023history,sartori2012emg}. 
Platforms such as OpenSim and CEINMS have facilitated this shift by supporting modular implementations and flexible parameter calibration~\cite{delp2007opensim,modenese2016estimation,pizzolato2015ceinms,sartori2014hybrid}.

However, a fundamental limitation persists: most EMG-driven MSK models are calibrated under a single loading condition and assume that model parameters remain invariant across different external mechanical demands. 
This assumption does not reflect real-world or clinical scenarios, where many activities, such as lifting, reaching, or rehabilitation exercises, may appear visually similar but involve markedly different mechanical contexts due to varying external loads~\cite{crossley2025calibrated,van2022building}. 
Failing to account for these load-dependent effects may compromise model accuracy and reduce generalizability.

Emerging studies increasingly challenge this assumption. Crossley et al.~\cite{crossley2025calibrated} showed that load-specific calibration significantly improves joint force predictions. 
Moreover, physiological evidence suggests that neuromuscular parameters such as EMG-to-activation dynamics, muscle force-generating capacity, and musculotendon geometry can adapt with changes in external load~\cite{roman1993adaptations,cheng2020load,kruse2021stimuli}. 
Thus, there is a critical need to investigate how MSK parameters vary with load and to establish calibration strategies that account for these effects. 
Despite isolated efforts in load-specific modeling, a systematic, quantitative analysis of load-aware EMG-driven calibration for upper-limb MSK models remains lacking.

In this study, we position load-aware calibration as the central focus: we aim to determine how external loading influences musculoskeletal parameter estimation and joint torque prediction accuracy. 
We hypothesize that MSK parameters are load-sensitive and that explicitly accounting for external load during calibration improves the model's ability to distinguish between kinematically similar but mechanically distinct tasks.
To support this investigation, we further evaluate
(1) the role of optimization frameworks in producing stable and physiologically plausible parameter estimates across load conditions and 
(2) the sensitivity of individual parameters in affecting torque prediction to identify those most critical for effective load-aware modeling.

A number of global and hybrid optimization techniques, including simulated annealing (SA), particle swarm optimization (PSO), and PSO combined with pattern search (PSO-PS), have been previously used for MSK calibration~\cite{pizzolato2015ceinms, schutte2005evaluation, jiang2025personalized}. 
However, how these frameworks perform under variable loading contexts, particularly in EMG-driven models, remains underexplored.
Additionally, studies have shown that only a few key parameters dominate MSK model sensitivity~\cite{reed2015optimising,hinson2022sensitivity,hosseini2022uncertainty,li2025parameter,arroyave2020multivariate}. 
Understanding which parameters are most load-sensitive can improve calibration efficiency and robustness, especially when data is limited.

Taken together, this work aims to provide a principled evaluation of load-aware calibration for EMG-driven MSK modeling, while investigating supporting aspects such as optimization consistency and parameter sensitivity. 
By enabling models to accurately distinguish between mechanically distinct tasks despite visual similarity, this study addresses a critical barrier to robust, real-world deployment of MSK frameworks.
\subsection*{Study Objectives and Contributions}

This study aims to evaluate how external mechanical loading influences the calibration and performance of EMG-driven musculoskeletal (MSK) models of the elbow joint. We hypothesize that MSK parameters are load-sensitive and that explicitly incorporating load-specific calibration improves joint torque prediction and model generalizability. To this end, our key contributions are as follows:
\begin{itemize}
    \item Demonstrate that load-aware, subject-specific calibration improves torque estimation by enabling the model to distinguish between visually similar but mechanically distinct tasks. This is validated across three hand load conditions (0, 2, and 4\,kg) using load-specific, cross-load, and global calibration strategies.
    
    \item Benchmark three global optimization frameworks, Simulated Annealing (SA), Particle Swarm Optimization (PSO), and a hybrid PSO-pattern search (PSO-PS), to assess their impact on the stability and physiological plausibility of the calibrated MSK parameters.
    
    \item Quantify the sensitivity of individual musculoskeletal parameters by analyzing their influence on torque prediction accuracy using root mean squared error (RMSE) and Pearson correlation metrics, thereby identifying parameters critical for robust and efficient calibration.
\end{itemize}
The proposed load-aware EMG-driven musculoskeletal (MSK) modeling framework is validated in a controlled experimental setting using data from eleven subjects performing elbow flexion-extension movements under varying external hand loads (0, 2, and 4~kg). 
This experimental design is chosen to mimic real-world scenarios where the same movement pattern may involve different mechanical demands. 
Model-predicted joint torques, generated from calibrated EMG-driven parameters, are directly compared with experimental torques estimated via subject-specific inverse dynamics.

The remainder of this article is organized as follows: Section~\ref{sec:background} summarizes the modeling background and problem formulation. Section~\ref{sec:methods} details the experimental protocol, data processing, model implementation, and calibration strategies. Section~\ref{sec:results} presents key findings, and Section~\ref{sec:discussion} discusses their implications. Section~\ref{sec:conclusion} concludes the paper.
\section{Background and Problem Formulation}
\label{sec:background}

\subsection{General EMG-Driven Musculoskeletal Modeling}
\label{subsec:emgdriven_general}

Electromyography (EMG)-driven musculoskeletal (MSK) models have become increasingly popular for non-invasively estimating muscle forces and joint torques in vivo~\cite{Buchanan2004neuromusculoskeletal, sartori2012emg, zajac1989muscle}. These models establish a physiologically grounded mapping from recorded EMG signals to joint biomechanics through four key stages:

\begin{itemize}
    \item \textbf{EMG-to-activation dynamics:} Pre-processed EMG envelopes $e_j[n]$ are transformed into neural excitations $u_j[n]$ using a second-order recursive filter:
    \begin{equation}\label{eq:emg_to_excitation}
        u_j[n] = \alpha\, e_j[n-d] - \beta_1\, u_j[n-1] - \beta_2\, u_j[n-2]
    \end{equation}
    where $\alpha$, $\beta_1$, $\beta_2$ are filter coefficients, and $d$ is the electromechanical delay. The resulting excitation is passed through a nonlinear function to yield muscle activation $a_j[n]$~\cite{lloyd2003emg, winby2009muscle}:
    \begin{equation}\label{eq:excitation_to_activation}
        a_j[n] = \frac{e^{A_j u_j[n]} - 1}{e^{A_j} - 1}
    \end{equation}
    where $A_j$ determines the nonlinearity.
    
    \item \textbf{Musculotendon kinematics:} Muscle fiber length $L_{mj}[n]$ and moment arm $M_{aj}[n]$ are modeled as functions of joint angles $q_i[n]$:
    \begin{align}
        L_{mj}[n] &= s_{L_{mj}} \left( \kappa_j + \sum_{i=1}^{2} y_{i}\, q_i[n] \right) 
        \label{eq:muscle_length_general}\\
        M_{aj}[n] &= s_{M_{aj}} \left( r_j + \sum_{i=1}^{2} x_{i}\, q_i[n] \right)
        \label{eq:moment_arm_general}
    \end{align}
    where $s_{L_{mj}}$, $s_{M_{aj}}$ are scaling parameters.
    
    \item \textbf{Muscle contraction dynamics:} Total muscle force $F_j^{m}[n]$ is estimated using a Hill-type formulation:
    \begin{equation}
        F_j^{m}[n] = \left[ F_j^{\mathrm{CE}}[n] + F_j^{\mathrm{PE}}[n] \right] \cos(\phi_j[n])
        \label{eq:muscle_force_general}
    \end{equation}
    where $F_j^{\mathrm{CE}}$ and $F_j^{\mathrm{PE}}$ are contractile and passive elements, respectively. Contractile force is:
    \begin{equation}
        F_j^{\mathrm{CE}}[n] = F_{oj}^{m}\, a_j[n]\, \bar{f}_L^{a}(\bar{l}_j^{m}[n])\, \bar{f}_V(\bar{v}_j^{m}[n])
    \end{equation}
    and passive force is:
    \begin{equation}
        F_j^{\mathrm{PE}}[n] = F_{oj}^{m}\, \bar{f}_L^{p}(\bar{l}_j^{m}[n])
    \end{equation}
    where $F_{oj}^{m}$ is the maximum isometric force, $\bar{f}_L^{a}(\cdot)$, $\bar{f}_V(\cdot)$, and $\bar{f}_L^{p}(\cdot)$ are normalized active force-length, force-velocity, and passive force-length relationships, respectively (see Appendix D\eqref{app:contraction-details}) and $\bar{l}_j^{m}[n]$, $\bar{v}_j^{m}[n]$ are the normalized fiber length and velocity, respectively.

    The pennation angle $\phi_j[n]$ is computed as:
    \begin{equation}
        P_{aj}[n] = 
        \begin{cases}
            0, & L_{mj}[n] = 0 \text{ or } w_j[n] \leq 0 \\
            \sin^{-1}\big(w_j[n]\big), & 0 < w_j[n] < 1 \\
            \frac{\pi}{2}, & w_j[n] \geq 1
        \end{cases}
    \end{equation}
    where $w_j[n] = \frac{L_{mj}^{o} \sin(P_{aj}^{o})}{L_{mj}[n]}$.
    
    \item \textbf{Joint torque estimation:} Net torque at each joint is computed as the sum of muscle moments:
    \begin{equation}\label{eq:joint_torque_general}
        \tau[n] = \sum_{j=1}^{k} F_j^{m}[n]\, M_{aj}[n]
    \end{equation}
\end{itemize}

While these models offer powerful capabilities, most studies treat the involved parameters (e.g., activation dynamics, force scaling, geometry) as fixed per subject and task, and calibrate them by minimizing the error between predicted and measured torques~\cite{winby2009muscle,lloyd2003emg}. This conventional approach implicitly assumes that the underlying biomechanical parameters remain invariant, even when external mechanical loading varies. As such, it fails to capture task-specific physiological adaptation, particularly in situations where activities are visually similar but mechanically distinct.

\subsection{Motivation for Load-Aware Calibration}
\label{subsec:motivation_load}

Physiological and biomechanical studies have shown that external mechanical load can alter muscle behavior through neural, muscular, and structural adaptation~\cite{roman1993adaptations,cheng2020load,kruse2021stimuli}. Consequently, model parameters such as maximum isometric force, optimal fiber length, EMG-to-activation filter dynamics, and moment arms may vary across loading conditions. Nevertheless, traditional EMG-driven MSK models ignore this variation, assuming parameter consistency regardless of load.

This gap is particularly problematic for tasks that are kinematically similar but differ in their mechanical demands. As a result, a key question arises: Can MSK models be made more accurate and generalizable by explicitly incorporating load-specific parameter calibration?

We formulate the following guiding questions:
\begin{itemize}
    \item How do EMG-driven MSK parameters vary across different external loads?
    \item Which parameters are most sensitive to loading changes?
    \item Can load-aware calibration improve joint torque estimation and enable generalization across tasks with different mechanical profiles?
\end{itemize}

\subsection{Problem Formulation: Load-Aware EMG-Driven Calibration}
\label{subsec:problem_loadaware}

To answer these questions, we propose a load-aware calibration framework for EMG-driven MSK modeling of elbow flexion-extension. Our goal is to calibrate MSK parameters under multiple external load conditions and systematically assess their variation and generalizability.
Given synchronized EMG and joint kinematic data from a subject performing elbow motion under hand loads of 0, 2, and 4\,kg, our framework proceeds as follows:
\begin{enumerate}
    \item Build a subject-specific MSK model assuming stiff tendons and the dynamic formulation described above.
    \item Independently calibrate muscle-specific parameters for each load condition using inverse dynamics torque data.
    \item Quantify parameter variability across loads and identify load-sensitive parameters.
    \item Assess cross-load generalizability by applying calibrated parameters from one load to other load conditions.
\end{enumerate}
Mathematically, the calibration problem for each load $l$ is formulated as minimizing the squared error between the model-predicted torque and experimental (inverse dynamics) torque:
\begin{equation}
    \min_{\boldsymbol{p}^{(l)}} \mathcal{L}^{(l)} = \frac{1}{T} \sum_{n=1}^T \left( \tau_{\text{model}}^{(l)}(n;\boldsymbol{p}^{(l)}) - \tau_{\text{exp}}^{(l)}[n] \right)^2
    \label{eq:optimization_loss}
\end{equation}
subject to physiological constraints on the parameter set:
{\footnotesize
\begin{align*}
    & c_1^{(l)},\, c_2^{(l)} \in [-1, 1];\quad
      A_j^{(l)} \in [-3, 0];\quad
      F_{oj}^{m,(l)} \in [F_{oj,\text{min}}^{m}, F_{oj,\text{max}}^{m}]; \\
    & L_{mj}^{o,(l)} \in [L_{mj,\text{min}}^{o}, L_{mj,\text{max}}^{o}];\quad
      P_{aj}^{o,(l)} \in [P_{aj,\text{min}}^{o}, P_{aj,\text{max}}^{o}]; \\
    & s_{L_{mj}}^{(l)} \in [s_{L_{mj,\text{min}}}, s_{L_{mj,\text{max}}}];\quad
      s_{M_{aj}}^{(l)} \in [s_{M_{aj,\text{min}}}, s_{M_{aj,\text{max}}}],\quad 
      \forall j = 1,\ldots,N.
\end{align*}
}

Here, $\boldsymbol{p}^{(l)} = \{c_1, c_2, A_j, F_{oj}^{m}, L_{mj}^{o}, P_{aj}^{o}, s_{L_{mj}}, s_{M_{aj}}\}_{j=1}^{N}$ represents the calibrated parameters for load $l$. Comparing parameter sets $\mathcal{P}^{(0)}, \mathcal{P}^{(2)}, \mathcal{P}^{(4)}$ allows us to quantify load-induced variation. Additionally, applying $\mathcal{P}^{(l)}$ to unseen loads ($l' \ne l$) provides insight into cross-load generalization, evaluated via RMSE and Pearson correlation between predicted and measured torques.
This load-aware formulation establishes a principled basis for studying the physiological sensitivity and robustness of EMG-driven MSK models under varying mechanical contexts, with direct implications for clinical, ergonomic, and assistive applications.
\begin{figure}[ht!]
\centerline{\includegraphics[width=\linewidth]{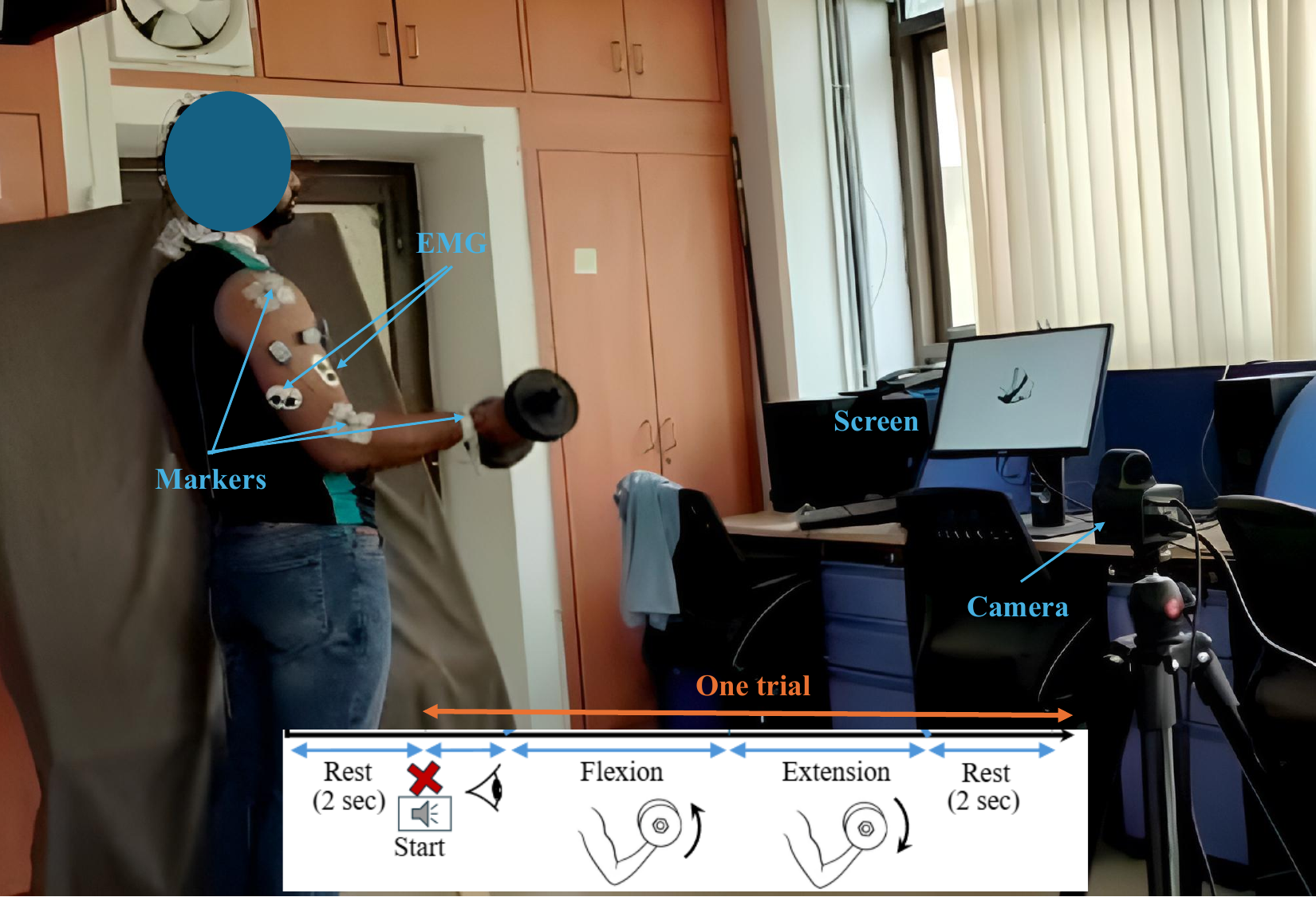}}
\caption{\textbf{Experimental data recording setup and protocol.}
(a) Surface EMG electrodes were placed on four upper-arm muscles (biceps long head, biceps short head, triceps long head, and triceps lateral head), and reflective markers were attached for joint angle measurement. 
Elbow flexion-extension movements were performed by participants while holding dumbbells of 0, 2, or 4 kg. EMG and joint kinematic data were acquired synchronously using a Noraxon Ultium wireless system and NiNOX camera, respectively.
(b) The experimental protocol comprised alternating phases of preparation, elbow flexion-extension task, and rest, with visual and auditory cues provided via a display and speaker.}
\label{fig:experimental_setup}
\end{figure}
\section{Methods}
\label{sec:methods}

\subsection{Participants and experimental protocol}
\label{subsec:participants}
Eleven healthy, right-handed adult volunteers participated in this study (see Table~\ref{table:group_demographics}). All experimental procedures were approved by the Institutional Review Board of the All India Institute of Medical Sciences, New Delhi, with informed consent obtained from each participant. 
Surface EMG was recorded from four upper-arm muscles spanning agonist–antagonist pairs (biceps long head/short head, triceps long head/lateral head; Fig.~\ref{fig:experimental_setup}), capturing variations in muscle coordination due to different mechanical loadings.
EMG electrode placement and skin preparation followed standard guidelines~\cite{konrad2005abc}. 
EMG signals were recorded using Noraxon’s wireless Ultium EMG sensor system at a sampling rate of 4000 Hz. The elbow joint angle was measured using a marker-based camera system (Noraxon NiNOX 125) positioned two meters away from the subject in the sagittal plane. 
The camera system was integrated with the Noraxon myoResearch platform (MR 3.16) to capture elbow flexion-extension (eFE) movements. During post-processing, the elbow joint angle was determined using the myoResearch software. 
Reflective markers were tracked with a three-point angle measurement tool to compute the 2-D joint angle from the video recordings. The joint angle data were sampled at 125 Hz, and synchronization of the sEMG and joint angle data was achieved using the Noraxon myoSync device.
\begin{table}[h!]
    \centering
    \caption{Summary of demographic characteristics (mean~$\pm$~SD) for each subject group.}
    \begin{adjustbox}{width=0.485\textwidth}
    \begin{tabular}{l l ccc}
    \hline
    Group & Subject IDs & Age (years) & Weight (kg) & Height (cm) \\ \hline
    Young Males (n=4) & S-1 to S-4 & 26.4$\pm$2.0 & 73.2$\pm$3.1 & 175.1$\pm$3.0 \\
    Young Females (n=3) & S-5 to S-7 & 28.7$\pm$1.5 & 58.3$\pm$6.0 & 154.8$\pm$5.0 \\
    Middle-aged Males (n=4) & S-8 to S-11 & 44.0$\pm$6.5 & 66.8$\pm$17.9 & 167.0$\pm$5.7 \\ \hline
    \end{tabular}
    \end{adjustbox}
    \label{table:group_demographics}
\end{table}
\vspace{-2em}
\subsection{Data acquisition}
\label{subsec:data_acquisition}
During each experimental session, both surface EMG and joint kinematic data were recorded as participants performed repeated elbow flexion–extension (eFE) tasks while holding dumbbells of 0, 2, or 4 kg in their right hand. 
Experimental instructions were presented on a monitor placed two meters ahead of the participant, and the protocol was managed using PsychoPy \cite{peirce2007psychopy}. 
Each trial began with a preparatory cross and auditory beep, followed by a visual cue signaling the start of the eFE movement. After completing the flexion–extension movement, participants held a standardized posture for a two-second resting phase. 
Prior to the main trials, participants received verbal instructions and practiced the task to ensure familiarity.
The experiment comprised six sets per subject, two sets for each load condition (0, 2, and 4~kg), with each set containing ten eFE task trials. Rest periods were scheduled between sets to minimize muscle fatigue. 
For model development, one set per load was used for calibration and the other for testing, enabling evaluation of both within-load and cross-load generalizability.
Notably, this protocol produced kinematically similar but mechanically distinct tasks, providing an ideal testbed for evaluating load-aware musculoskeletal model calibration.
\subsection{Signal processing and data preprocessing}
\label{subsec:signal_processing}
Raw EMG signals were preprocessed to improve signal quality and extract physiologically relevant features. 
Signals were band-pass filtered (10–450 Hz, 4th-order Butterworth) to remove motion artifacts and high-frequency noise, then full-wave rectified and low-pass filtered at 7 Hz to obtain the linear envelope~\cite{sartori2012emg}. 
EMG envelopes were downsampled to 125~Hz to match the kinematic data and normalized to each subject’s maximum voluntary contraction (MVC), determined in separate trials.
Elbow joint angles were extracted from marker-based motion capture, then denoised with a Gaussian filter (window size 6, $\sigma$=10). 
Each eFE task was segmented into individual cycles using joint velocity profiles and resampled to a fixed length, allowing direct averaging and comparison of trials across subjects and conditions.

\subsection{Inverse dynamics torque estimation}
To obtain the ground-truth elbow joint torques required for musculoskeletal model calibration, we employed a subject-specific inverse dynamics approach using OpenSim’s \textit{arm26.osim} model \cite{delp2007opensim}. 
For each participant, the generic \textit{arm26.osim} musculoskeletal model was scaled to match individual anthropometry based on reflective marker positions captured during static and dynamic calibration trials. 
Scaling and kinematic fitting were performed using the OpenSim GUI and API tools.
The external loading condition was modeled by attaching a virtual dumbbell (Fig.~\ref{fig:experimental_setup}) to the distal end of the forearm segment. The dumbbell mass (0 kg, 2 kg, or 4 kg) was configured as an external force applied at the hand marker location, simulating the added load during elbow flexion-extension tasks. 
Each load condition corresponded to a distinct experimental protocol, with subjects performing identical movements under the three load levels.
Time-varying joint angles were derived from the marker trajectories and used to drive the scaled musculoskeletal model. 
Using the OpenSim inverse dynamics tool, the net joint torques were computed at the elbow by solving the Newton-Euler equations of motion, accounting for segmental inertia, gravitational effects, and the modeled external load. The resulting load-specific torque profiles served as the reference for validating load-aware EMG-driven models.

\subsection{Musculoskeletal model implementation}
\label{subsec:model_implementation}
The EMG-driven musculoskeletal (MSK) model, as detailed in Section~\eqref{sec:background}, was implemented in MATLAB (MathWorks Inc., R2024b). 
The model adopts a standard Hill-type muscle representation under the stiff tendon assumption, which isolates the contractile dynamics for improved identifiability.
For each muscle, processed EMG envelopes were converted to neural activation via a second-order recursive filter and a nonlinear static mapping (see Eqns.~\eqref{eq:emg_to_excitation}--\eqref{eq:excitation_to_activation}). 
Muscle–tendon unit lengths and moment arms were computed as subject- and load-specific polynomial functions of joint angle (Eqns.~\eqref{eq:muscle_length_general}--\eqref{eq:moment_arm_general}), and muscle forces were determined using established force-length and force-velocity relationships (Eqns.~\eqref{eq:muscle_force_general}). The net elbow joint torque was calculated as the sum of muscle moments (Eqn.~\eqref{eq:joint_torque_general}).
\begin{figure}[ht!]
\centerline{\includegraphics[width=\linewidth]
{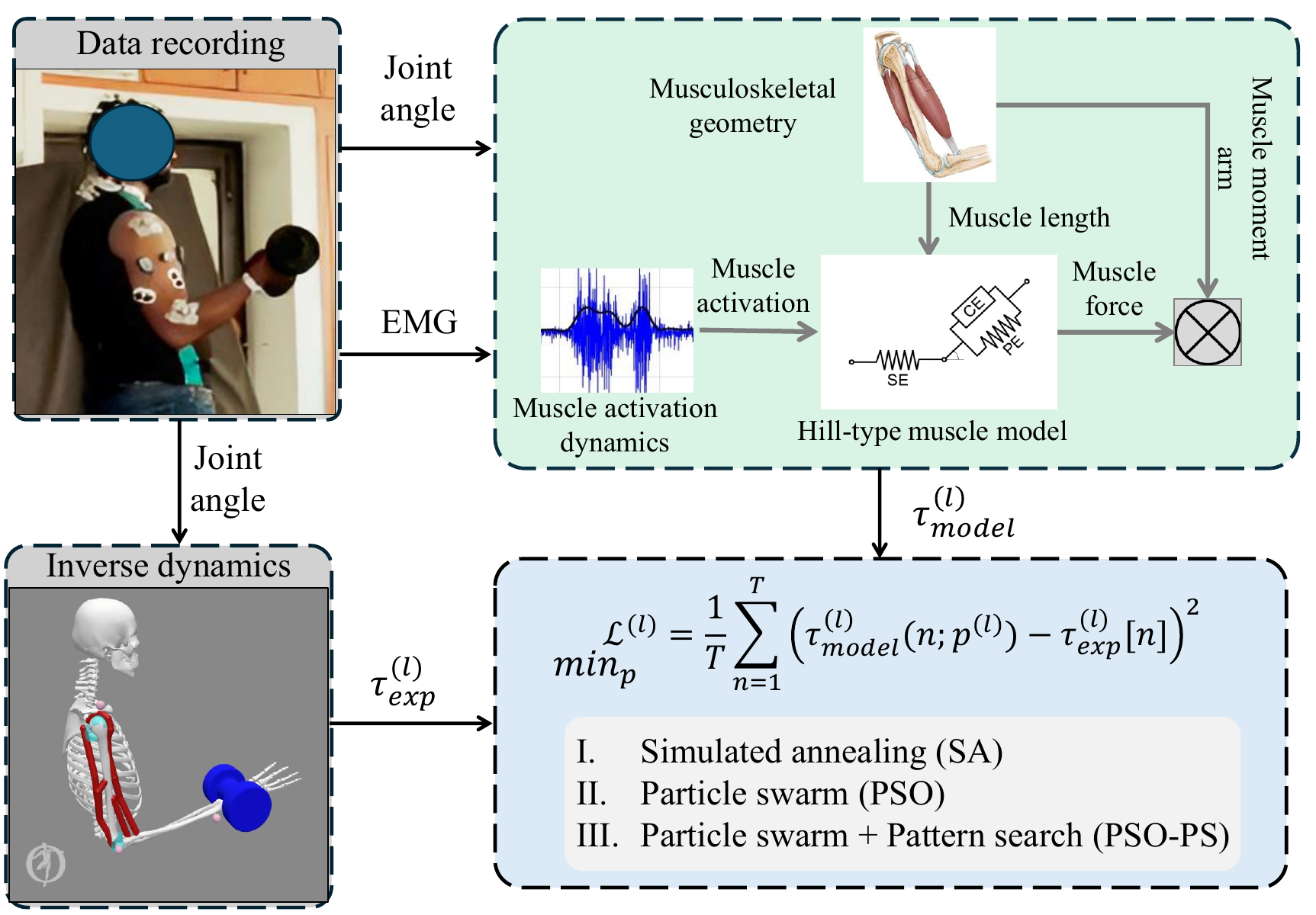}}
\caption{\textbf{Main framework.} This modeling setup was designed to isolate parameter responses to load variations while ensuring consistency in movement kinematics.}
\label{fig:main_framework}
\end{figure}
\subsection{Parameter calibration and optimization}
\label{subsec:calibration}
The calibration objective was to identify, for each subject under each external load condition ($l \in {0, 2, 4}$kg), a load-specific set of muscle model parameters that best reproduced the experimentally measured joint torques, given the subject’s EMG and joint kinematic data. 
This load-aware calibration approach allows the model to explicitly account for physiological and mechanical variations induced by different loading conditions. 
The calibrated parameters included scaling factors for maximum isometric force ($s_{F^{m}{o}}^{(l)}$), optimal fiber length ($s{L^{m}{o}}^{(l)}$), pennation angle ($s{P_{a}^{o}}^{(l)}$), EMG-to-activation dynamics ($c_1^{(l)}, c_2^{(l)}, A^{(l)}$), and musculotendon geometry ($s_{L_{m}}^{(l)}, s_{M_{a}}^{(l)}$), as detailed in Table \ref{tab:calibrated_parameters}. 
All parameters were constrained within physiologically plausible bounds to ensure realistic muscle behavior.
\begin{table}[ht]
\centering
\caption{Calibrated model parameters, physiological meaning, and bounds (per muscle and load).}
\label{tab:calibrated_parameters}
\begin{adjustbox}{width=0.49\textwidth}
\begin{tabular}{ccc}
\toprule
Parameter & Meaning & Bound/Range \\
\midrule
$s_{F^{m}_{o}}^{(l)}$ & Max isometric force scaling & [0.5, 2.5] \\
$s_{L^{m}_{o}}^{(l)}$ & Optimal fiber length scaling & [0.95, 1.05] \\
$s_{P_{a}^{o}}^{(l)}$ & Pennation angle scaling & [0.95, 1.05] \\
$c_1^{(l)}, c_2^{(l)}$ & EMG-to-activation filter & [-1, 1] \\
$A^{(l)}$ & Nonlinearity shape factor & [-3, 0] \\
$s_{L_{m}}^{(l)}$ & Muscle length polynomial scale & [0.5, 1.5] \\
$s_{M_{a}}^{(l)}$ & Moment arm polynomial scale & [0.5, 1.5] \\
\bottomrule
\end{tabular}
\end{adjustbox}
\end{table}

\subsubsection{Optimization framework}

The calibration task was formulated as a nonlinear optimization problem, where the goal was to determine a load-specific parameter set $\boldsymbol{p}^{(l)}$ for each subject and load condition ($l \in \{0, 2, 4\}$~kg) that minimizes the mean squared error between model-predicted and experimentally measured joint torques (see Eqn.~\ref{eq:optimization_loss}). This formulation enables the explicit capture of load-induced physiological variability.
To ensure robust convergence, we employed three optimization frameworks:
\begin{itemize}
    \item \textbf{Simulated Annealing (SA):} A stochastic global optimization method effective in escaping local minima.
    \item \textbf{Particle Swarm Optimization (PSO):} A population-based global search algorithm that simulates collective behavior.
    \item \textbf{Hybrid PSO-pattern search (PSO-PS):} A two-stage method combining PSO for global exploration and pattern search for local refinement.
\end{itemize}
All optimization procedures were implemented using MATLAB’s Global Optimization Toolbox, with physiological bounds enforced on all parameters (see Appendix C \eqref{app:main_pseudocode}\& \eqref{app:optimizer_pseudocode}). Optimizer-specific settings are listed in Table~\ref{tab:optimizer_settings}.
\begin{equation}
\min_{\boldsymbol{p}^{(l)}} \mathcal{L}^{(l)} = \frac{1}{T} \sum_{n=1}^T \left( \tau_{\text{model}}^{(l)}(n;\boldsymbol{p}^{(l)}) - \tau_{\text{exp}}^{(l)}[n] \right)^2
\label{eq:optimization_loss}
\end{equation}
\begin{table}[ht]
\centering
\caption{Parameter settings for each optimization method.}
\label{tab:optimizer_settings}
\begin{tabular}{ccc}
\toprule
\textbf{Optimizer} & \textbf{Parameter} & \textbf{Value} \\
\midrule
\multirow{2}{*}{SA} & MaxIterations & 5000 \\
                    & ObjectiveLimit & 0.5 \\
\midrule
\multirow{2}{*}{PSO} & SwarmSize & 30 \\
                     & MaxIterations & 250 \\
\midrule
\multirow{3}{*}{PSO-PS} & PSO SwarmSize & 30 \\
                        & PSO MaxIterations & 250 \\
                        & Pattern Search MaxIter & 2500 \\
\bottomrule
\end{tabular}
\end{table}
\subsubsection{Calibration Strategies}

To systematically assess the effectiveness of load-aware calibration, we evaluated three primary calibration strategies:

\begin{itemize}
    \item \textbf{Load-Specific (Load-Aware):} Parameters were independently optimized for each load condition (0, 2, and 4~kg) to capture load-dependent physiological adaptations.
    \item \textbf{Global:} A single parameter set was calibrated using aggregated data from all load conditions, assuming load-invariant muscle behavior.
    \item \textbf{Cross-Load:} Parameters calibrated under one load (e.g., 0~kg) were applied to the remaining conditions (2~kg, 4~kg, and global) to assess generalization across mismatched loading contexts.
\end{itemize}

The five total configurations derived from these strategies are summarized in Table~\ref{tab:calibration_strategies}. In addition to prediction performance, we evaluated parameter robustness by quantifying inter-load variability using the coefficient of variation (CV) and percent deviation metrics.

\begin{table}[ht]
\centering
\caption{Calibration Strategies Evaluated}
\label{tab:calibration_strategies}
\begin{tabular}{p{2.5cm} p{5cm}}
\toprule
\textbf{Strategy} & \textbf{Description} \\
\midrule
Load-Specific & Independent optimization per load (0, 2, 4~kg) \\
Global & Single optimization using all load data \\
Fixed-0kg & 0~kg parameters applied to 2~kg, 4~kg, and global \\
Fixed-2kg & 2~kg parameters applied to other loads \\
Fixed-4kg & 4~kg parameters applied to other loads \\
\bottomrule
\end{tabular}
\end{table}
\subsection{Model evaluation metrics}
\label{subsec:evaluation}

Model performance and parameter robustness were assessed using a combination of quantitative accuracy metrics and statistical analyses:

\begin{itemize}
    \item \textbf{Root Mean Squared Error (RMSE):}  
    RMSE quantifies the average magnitude of error between model-predicted ($\hat{y}_i$) and experimentally measured ($y_i$) joint torques. It provides an absolute measure of deviation in Nm:
    \begin{equation}
    \text{RMSE} = \sqrt{ \frac{1}{N} \sum_{i=1}^{N} (y_i - \hat{y}_i)^2 }
    \end{equation}
    where $N$ is the number of time samples.

    \item \textbf{Pearson Correlation Coefficient ($r$):}  
    The Pearson correlation coefficient evaluates the temporal agreement between predicted and experimental torque waveforms, independent of scale or magnitude:
    \begin{equation}
        r = \frac{ \sum_{i=1}^{N} (y_i - \bar{y})(\hat{y}_i - \bar{\hat{y}}) }{ \sqrt{ \sum_{i=1}^{N} (y_i - \bar{y})^2 } \sqrt{ \sum_{i=1}^{N} (\hat{y}_i - \bar{\hat{y}})^2 } }
    \end{equation}
    where $\bar{y}$ and $\bar{\hat{y}}$ denote the mean of experimental and predicted signals, respectively.

    \item \textbf{Coefficient of Variation (CV):}  
    CV quantifies inter-subject variability in model error (e.g., RMSE) and helps evaluate robustness across the population. It is computed as:
    \begin{equation}
        \text{CV} = \frac{\sigma}{\mu} \times 100\%
    \end{equation}
    where $\sigma$ and $\mu$ represent the standard deviation and mean of the metric across subjects.

    \item \textbf{Parameter Variability Metrics:}  
    To identify load-sensitive model parameters, we computed the percent deviation and CV of calibrated parameter values across load conditions. This analysis supports the interpretation of physiological adaptability and model generalization under variable loading.

    \item \textbf{Statistical Analysis:}  
    One-way ANOVA was used to test for significant differences in model performance (e.g., RMSE, $r$) and parameter distributions across loading conditions and optimization methods. Post-hoc comparisons were conducted where applicable, and 95\% confidence intervals were reported.
\end{itemize}

\section{Results}\label{sec:results}

\subsection{Evaluation of optimization frameworks}

\begin{figure}[ht!]
\centerline{\includegraphics[trim=0.12cm 8.0cm 0.2cm 9.1cm, clip=true,width=0.49\textwidth]{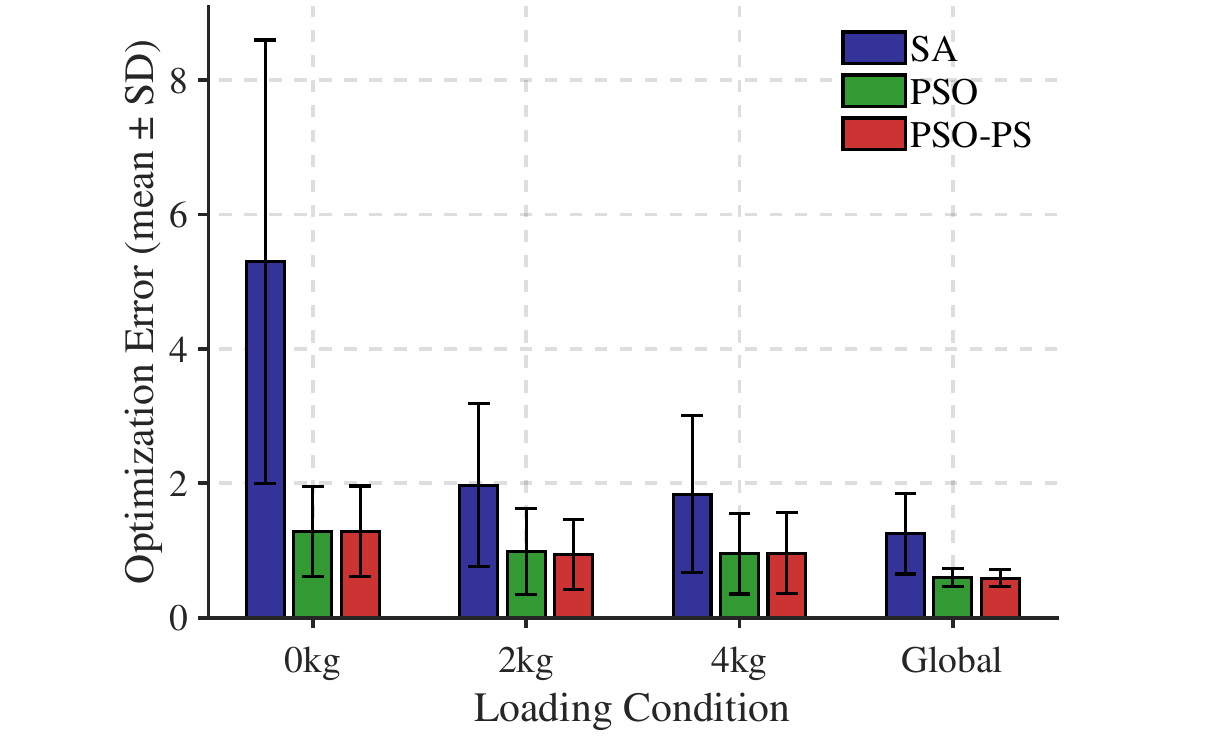}}
\caption{\textbf{Optimization method comparison for musculoskeletal calibration.}
Mean~$\pm$~SD optimization errors for simulated annealing (SA), particle swarm optimization (PSO), and hybrid PSO-pattern search (PSO-PS) across all subjects and loading conditions (0, 2, 4~kg, and global). PSO-PS consistently yields the lowest calibration error, highlighting its robustness for EMG-driven modeling under variable loads.}
\label{fig:optimization_error_methods}
\end{figure}

To assess the impact of optimization algorithms on musculoskeletal (MSK) parameter calibration, we compared SA, PSO, and a hybrid PSO-PS across all subjects and load conditions (0, 2, 4~kg, and global). 
As shown in Fig.~\ref{fig:optimization_error_methods}, PSO and PSO-PS consistently yielded lower mean torque estimation errors compared to SA, with this trend being more prominent under higher loading conditions. The hybrid PSO-PS approach produced the most stable outcomes across subjects and loads, likely due to its combination of global and local search mechanisms. 
However, PSO exhibited nearly equivalent performance to PSO-PS in terms of RMSE and correlation metrics, with marginal differences observed across most cases. These findings suggest that both population-based methods resulted in more consistent parameter estimation and torque prediction accuracy than SA under varying mechanical load conditions.
\begin{figure}[ht!]
\centerline{\includegraphics[trim=0cm 10cm 0cm 10cm, clip=true,width=\linewidth]{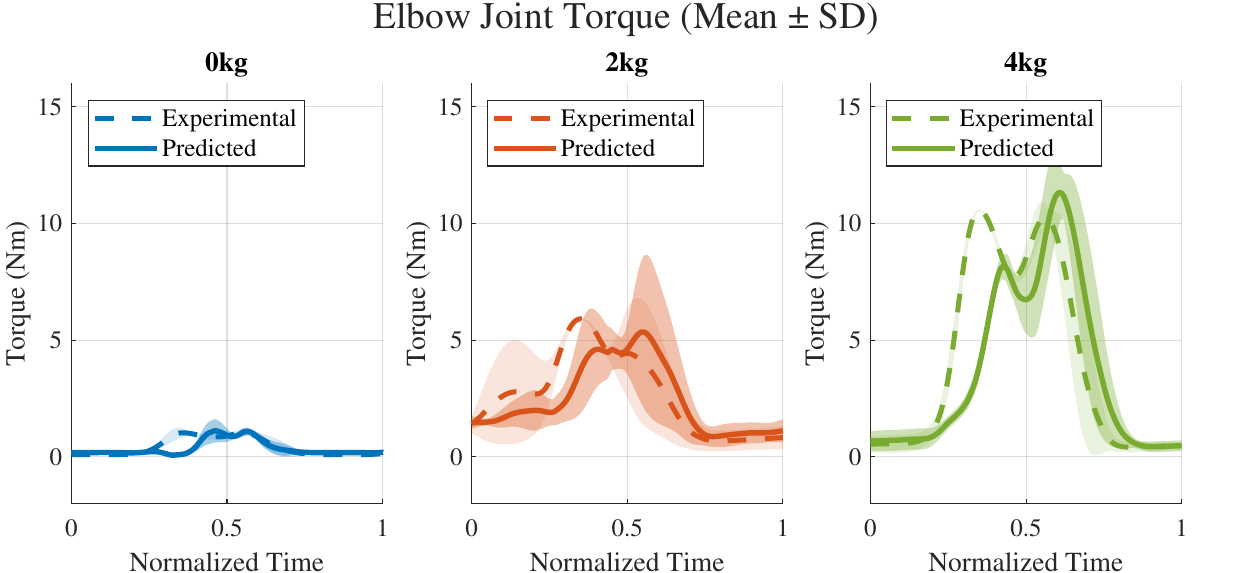}}
\caption{\textbf{Predicted vs. experimental elbow joint torque for subject S01.}
Mean (solid) and SD (shaded) for model-predicted and experimental torque across trials under all loads (0~kg: blue, 2~kg: red, 4~kg: green). Model predictions closely match experimental torque across conditions.}
\label{fig:torque_comparison}
\end{figure}
\begin{table*}[ht!]
\centering
\caption{RMSE (mean$\pm$std) and correlation (mean$\pm$std) for all subjects and load conditions. Correlation values with $p < 0.05$ are highlighted in green.}
\label{tab:subject_rmse_corr_color}
\renewcommand{\arraystretch}{1.15}
\setlength{\tabcolsep}{4pt}
\begin{adjustbox}{width=0.88\textwidth}
\begin{tabular}{
    >{\centering\arraybackslash}m{0.9cm}
    >{\centering\arraybackslash}m{0.8cm}
    >{\centering\arraybackslash}m{2cm}
    >{\centering\arraybackslash}m{2cm}
    >{\centering\arraybackslash}m{2cm}
    >{\centering\arraybackslash}m{2cm}
    >{\centering\arraybackslash}m{2cm}
    >{\centering\arraybackslash}m{2cm}
}
\toprule
\multirow{2}{*}{\textbf{Sub}} & \multirow{2}{*}{\textbf{Load}} 
& \multicolumn{3}{c}{\textbf{RMSE (mean $\pm$ std)}} 
& \multicolumn{3}{c}{\textbf{Correlation (mean $\pm$ std)}} \\
\cmidrule(lr){3-5} \cmidrule(lr){6-8}
& & \textbf{SA} & \textbf{PSO} & \textbf{PSO-PS} & \textbf{SA} & \textbf{PSO} & \textbf{PSO-PS} \\
\midrule
\multirow{3}{*}{S01} 
& 0kg & 1.02$\pm$0.15 & 0.46$\pm$0.02 & 0.46$\pm$0.02 
  & \cellcolor{sigcol}0.63$\pm$0.07 & \cellcolor{sigcol}0.56$\pm$0.06 & \cellcolor{sigcol}0.56$\pm$0.06 \\
& 2kg & 2.41$\pm$0.18 & 1.86$\pm$0.13 & 1.86$\pm$0.12 
  & \cellcolor{sigcol}0.54$\pm$0.03 & \cellcolor{sigcol}0.55$\pm$0.06 & \cellcolor{sigcol}0.53$\pm$0.06 \\
& 4kg & 3.97$\pm$0.27 & 3.51$\pm$0.25 & 3.51$\pm$0.25 
  & 0.39$\pm$0.06 & \cellcolor{sigcol}0.49$\pm$0.04 & \cellcolor{sigcol}0.49$\pm$0.04 \\
\midrule

\multirow{3}{*}{S02} 
& 0kg & 1.04$\pm$0.06 & 0.53$\pm$0.02 & 0.53$\pm$0.02 
  & 0.29$\pm$0.08 & \cellcolor{sigcol}0.52$\pm$0.04 & \cellcolor{sigcol}0.52$\pm$0.04 \\
& 2kg & 2.44$\pm$0.30 & 1.83$\pm$0.10 & 1.80$\pm$0.15 
  & \cellcolor{sigcol}0.65$\pm$0.02 & \cellcolor{sigcol}0.69$\pm$0.02 & \cellcolor{sigcol}0.71$\pm$0.04 \\
& 4kg & 4.16$\pm$0.29 & 2.99$\pm$0.22 & 2.99$\pm$0.22 
  & \cellcolor{sigcol}0.44$\pm$0.03 & \cellcolor{sigcol}0.68$\pm$0.04 & \cellcolor{sigcol}0.68$\pm$0.04 \\
\midrule

\multirow{3}{*}{S03}
& 0kg & 1.34$\pm$0.10 & 0.54$\pm$0.05 & 0.54$\pm$0.05
  & 0.41$\pm$0.09 & \cellcolor{sigcol}0.55$\pm$0.03 & \cellcolor{sigcol}0.55$\pm$0.03 \\
& 2kg & 2.53$\pm$0.05 & 1.75$\pm$0.05 & 1.80$\pm$0.06
  & 0.43$\pm$0.04 & \cellcolor{sigcol}0.64$\pm$0.02 & \cellcolor{sigcol}0.61$\pm$0.03 \\
& 4kg & 4.58$\pm$0.39 & 3.07$\pm$0.19 & 3.07$\pm$0.18
  & 0.42$\pm$0.07 & \cellcolor{sigcol}0.70$\pm$0.04 & \cellcolor{sigcol}0.70$\pm$0.04 \\
\midrule

\multirow{3}{*}{S04}
& 0kg & 0.59$\pm$0.08 & 0.36$\pm$0.02 & 0.33$\pm$0.01
  & 0.54$\pm$0.07 & \cellcolor{sigcol}0.74$\pm$0.01 & \cellcolor{sigcol}0.76$\pm$0.02 \\
& 2kg & 2.43$\pm$0.20 & 1.79$\pm$0.11 & 1.77$\pm$0.10
  & 0.40$\pm$0.11 & \cellcolor{sigcol}0.62$\pm$0.06 & \cellcolor{sigcol}0.63$\pm$0.06 \\
& 4kg & 3.99$\pm$0.62 & 2.89$\pm$0.42 & 2.89$\pm$0.42
  & 0.47$\pm$0.06 & \cellcolor{sigcol}0.67$\pm$0.05 & \cellcolor{sigcol}0.67$\pm$0.05 \\
\midrule

\multirow{3}{*}{S05}
& 0kg & 1.09$\pm$0.07 & 0.51$\pm$0.01 & 0.51$\pm$0.01
  & 0.22$\pm$0.04 & \cellcolor{sigcol}0.46$\pm$0.04 & \cellcolor{sigcol}0.46$\pm$0.04 \\
& 2kg & 3.25$\pm$0.83 & 1.81$\pm$0.52 & 1.81$\pm$0.52
  & 0.05$\pm$0.14 & \cellcolor{sigcol}0.52$\pm$0.08 & \cellcolor{sigcol}0.52$\pm$0.08 \\
& 4kg & 5.30$\pm$2.12 & 3.07$\pm$1.34 & 3.15$\pm$1.40
  & 0.15$\pm$0.05 & 0.44$\pm$0.10 & 0.41$\pm$0.12 \\
\midrule

\multirow{3}{*}{S06}
& 0kg & 0.79$\pm$0.05 & 0.43$\pm$0.03 & 0.43$\pm$0.02
  & 0.17$\pm$0.04 & 0.46$\pm$0.04 & 0.44$\pm$0.03 \\
& 2kg & 2.52$\pm$0.18 & 2.11$\pm$0.17 & 2.11$\pm$0.17
  & 0.24$\pm$0.12 & \cellcolor{sigcol}0.51$\pm$0.04 & \cellcolor{sigcol}0.51$\pm$0.06 \\
& 4kg & 4.21$\pm$0.65 & 3.09$\pm$0.41 & 3.10$\pm$0.41
  & 0.41$\pm$0.10 & \cellcolor{sigcol}0.56$\pm$0.07 & \cellcolor{sigcol}0.57$\pm$0.07 \\
\midrule

\multirow{3}{*}{S07}
& 0kg & 1.35$\pm$0.23 & 0.64$\pm$0.05 & 0.64$\pm$0.05
  & 0.12$\pm$0.10 & 0.17$\pm$0.09 & 0.17$\pm$0.09 \\
& 2kg & 2.95$\pm$0.75 & 2.05$\pm$0.77 & 2.04$\pm$0.76
  & 0.27$\pm$0.13 & 0.37$\pm$0.12 & 0.37$\pm$0.12 \\
& 4kg & 4.83$\pm$2.44 & 3.01$\pm$1.61 & 3.14$\pm$1.73
  & 0.39$\pm$0.15 & 0.43$\pm$0.15 & 0.43$\pm$0.14 \\
\midrule

\multirow{3}{*}{S08}
& 0kg & 0.63$\pm$0.11 & 0.32$\pm$0.10 & 0.32$\pm$0.10
  & 0.37$\pm$0.32 & 0.46$\pm$0.30 & 0.47$\pm$0.30 \\
& 2kg & 0.81$\pm$0.58 & 0.48$\pm$0.54 & 0.48$\pm$0.53
  & 0.40$\pm$0.30 & 0.49$\pm$0.30 & 0.49$\pm$0.30 \\
& 4kg & 4.37$\pm$1.54 & 3.31$\pm$1.24 & 3.31$\pm$1.23
  & 0.33$\pm$0.22 & 0.49$\pm$0.19 & 0.49$\pm$0.19 \\
\midrule

\multirow{3}{*}{S09}
& 0kg & 0.67$\pm$0.03 & 0.33$\pm$0.02 & 0.33$\pm$0.03
  & 0.37$\pm$0.05 & \cellcolor{sigcol}0.63$\pm$0.04 & \cellcolor{sigcol}0.62$\pm$0.05 \\
& 2kg & 2.23$\pm$0.13 & 1.42$\pm$0.13 & 1.42$\pm$0.13
  & \cellcolor{sigcol}0.56$\pm$0.03 & \cellcolor{sigcol}0.76$\pm$0.03 & \cellcolor{sigcol}0.76$\pm$0.03 \\
& 4kg & 3.26$\pm$0.11 & 2.65$\pm$0.14 & 2.65$\pm$0.14
  & \cellcolor{sigcol}0.70$\pm$0.02 & \cellcolor{sigcol}0.78$\pm$0.02 & \cellcolor{sigcol}0.78$\pm$0.02 \\
\midrule

\multirow{3}{*}{S10}
& 0kg & 0.96$\pm$0.06 & 0.63$\pm$0.04 & 0.62$\pm$0.04
  & 0.35$\pm$0.08 & 0.41$\pm$0.06 & 0.42$\pm$0.06 \\
& 2kg & 3.48$\pm$0.24 & 2.49$\pm$0.16 & 2.22$\pm$0.16
  & 0.05$\pm$0.18 & 0.17$\pm$0.24 & 0.20$\pm$0.27 \\
& 4kg & 4.60$\pm$0.90 & 3.39$\pm$0.80 & 3.24$\pm$0.98
  & $-$0.01$\pm$0.12 & 0.18$\pm$0.23 & 0.19$\pm$0.22 \\
\midrule

\multirow{3}{*}{S11}
& 0kg & 1.38$\pm$0.05 & 0.86$\pm$0.01 & 0.87$\pm$0.02
  & 0.32$\pm$0.04 & 0.37$\pm$0.03 & 0.34$\pm$0.03 \\
& 2kg & 4.19$\pm$0.18 & 2.43$\pm$0.10 & 2.43$\pm$0.10
  & 0.11$\pm$0.09 & \cellcolor{sigcol}0.45$\pm$0.11 & \cellcolor{sigcol}0.45$\pm$0.11 \\
& 4kg & 6.18$\pm$0.68 & 3.89$\pm$0.54 & 3.89$\pm$0.54
  & 0.20$\pm$0.04 & \cellcolor{sigcol}0.54$\pm$0.05 & \cellcolor{sigcol}0.54$\pm$0.05 \\
\bottomrule
\end{tabular}
\end{adjustbox}
\vspace{0.5ex}
\begin{minipage}{\textwidth}
\footnotesize
\textit{Note:} Correlation cells highlighted in green indicate statistical significance ($p < 0.05$) for the corresponding method/load/subject comparison. SA = Simulated Annealing, PSO = Particle Swarm Optimization, PSO-PS = Hybrid.
\end{minipage}
\end{table*}
\vspace{-1.2em}
\subsection{Evaluation of joint torque estimation}

To evaluate the predictive accuracy of the EMG-driven MSK model under different optimization frameworks, we compared the model-estimated joint torques with experimental measurements using RMSE and correlation coefficient ($r$) across all subjects, load conditions, and optimization methods (Table~\ref{tab:subject_rmse_corr_color}).
Both PSO and the hybrid PSO-PS frameworks exhibited comparable RMSE and correlation values across conditions, and generally outperformed SA, which showed higher RMSE and lower correlation, especially under higher load conditions. For example, for subject S01 at 2~kg, PSO and PSO-PS both yielded an RMSE of approximately 1.86~$\pm$~0.13~Nm with $r$ values above 0.53. Similarly, for S04 at 0~kg, the RMSE ranged between 0.33–0.36~Nm with $r$ values of 0.74–0.76 for both methods.
PSO-PS produced slightly more instances of statistically significant correlations ($p < 0.05$; highlighted in green) compared to PSO, indicating marginal improvements in capturing joint torque dynamics. Across all methods, model accuracy tended to improve with increasing external load, as reflected by lower RMSE and higher correlation values. For instance, for S09 at 4~kg, both PSO and PSO-PS achieved an RMSE of approximately 2.65~Nm with $r$ exceeding 0.78.
\begin{figure*}[ht!]
\centerline{\includegraphics[width=0.9\linewidth]{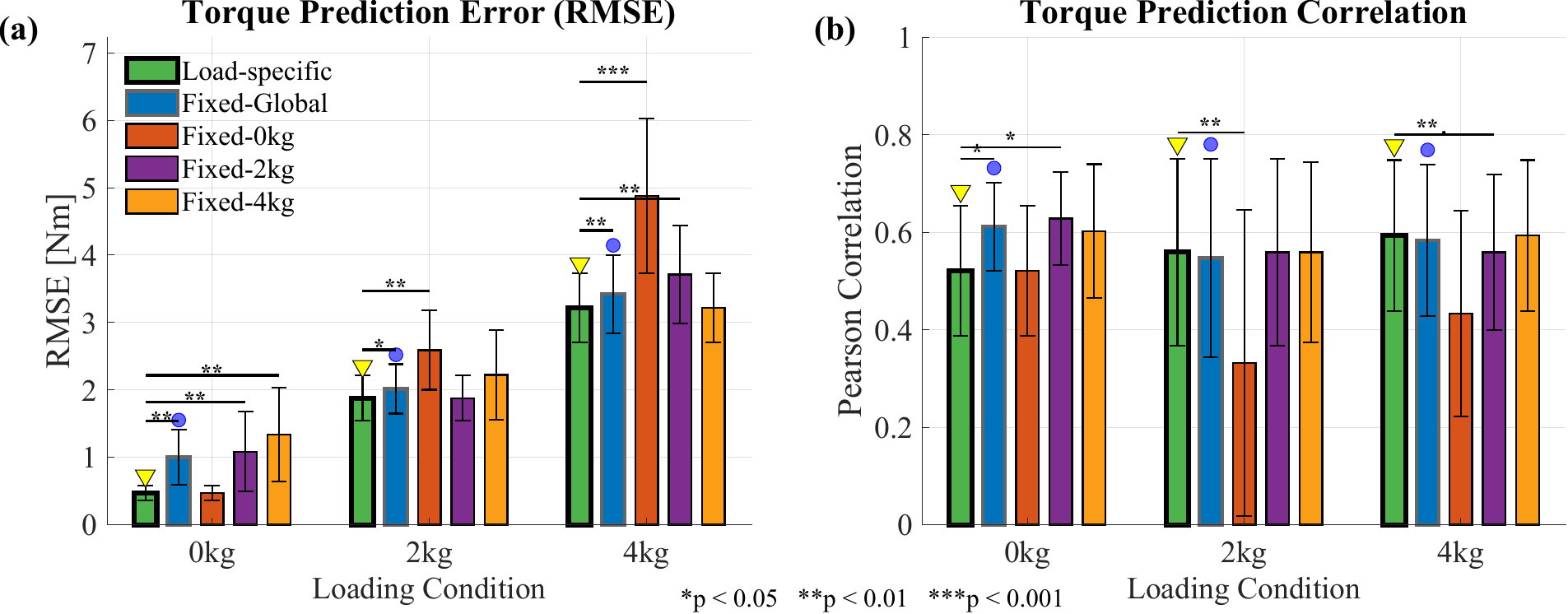}}
\caption{\textbf{Calibration strategy comparison for joint torque estimation across loading conditions.}
Mean~$\pm$~SD bar plots of (a) RMSE and (b) Pearson correlation ($r$) for five strategies—load-specific, fixed-global, fixed-0kg, fixed-2kg, and fixed-4kg—across all subjects and loads (0, 2, 4~kg). The top strategy (lowest RMSE, highest $r$) is marked with a black border/yellow triangle; the second-best with a gray border/blue circle.}
\label{fig:calibration_strategy}
\end{figure*}
\begin{figure}[ht!]
\centerline{\includegraphics[trim=0cm 10cm 0cm 10cm, clip=true,width=\linewidth]{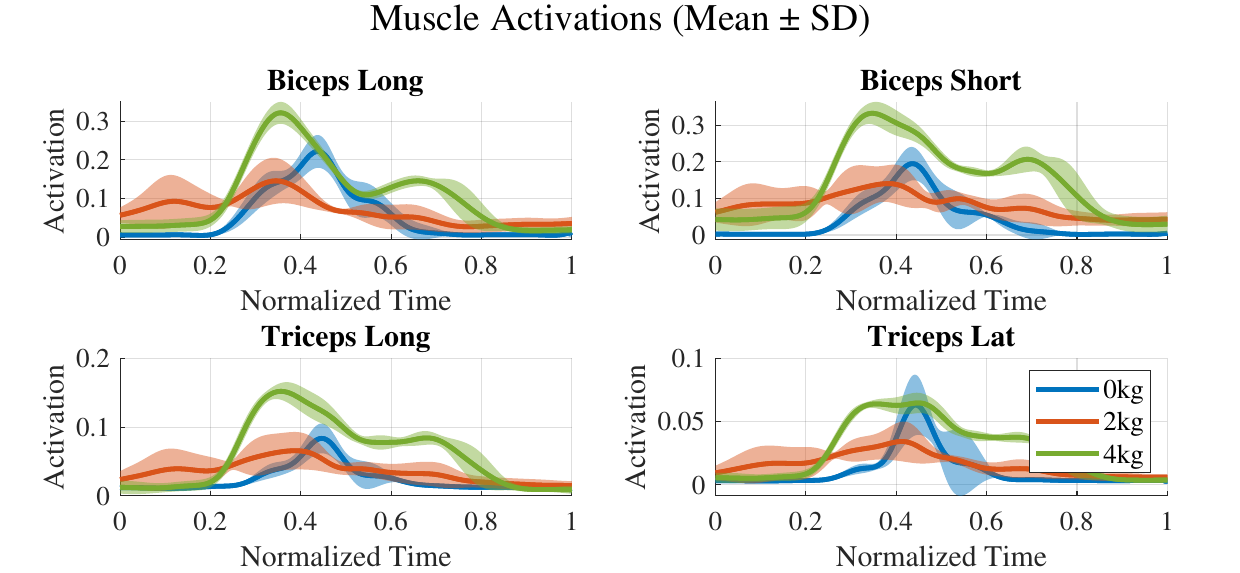}}
\caption{\textbf{Muscle activation trajectories (mean~$\pm$~SD) for biceps and triceps (subject S1).}
Average predicted activations for each muscle across trials and loads (colors: 0~kg blue, 2~kg red, 4~kg green). Shaded regions indicate trial-to-trial variability.}
\label{fig:activation_comparison}
\end{figure}
\begin{figure}[ht!]
\centerline{\includegraphics[trim=0cm 10cm 0cm 10cm, clip=true,width=\linewidth]{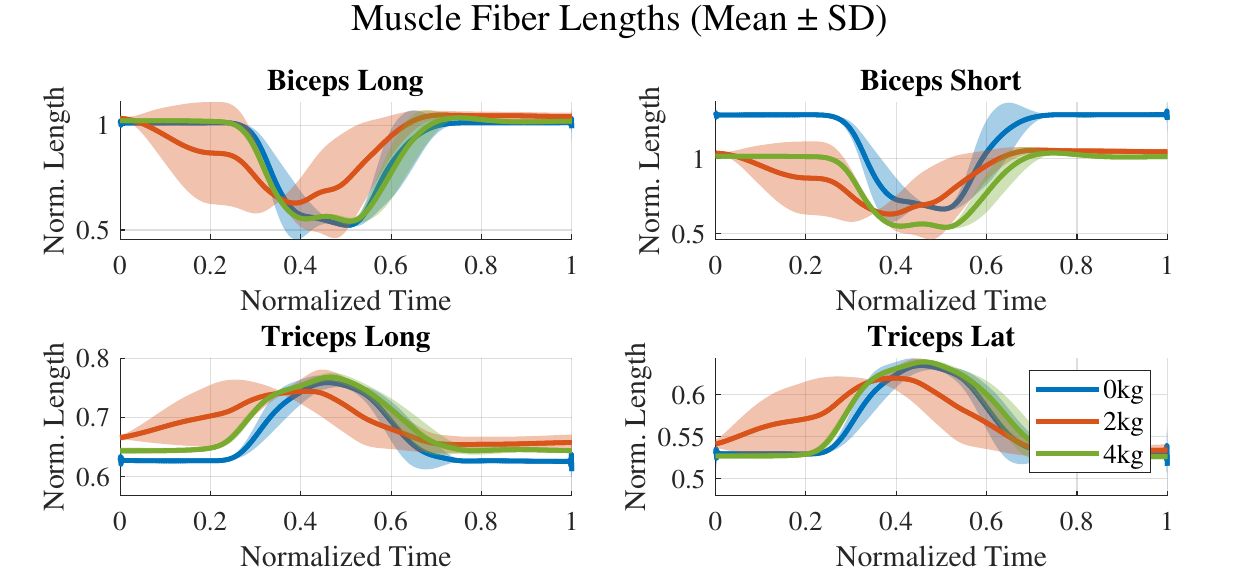}}
\caption{\textbf{Muscle fiber length (mean~$\pm$~SD) under different loads (subject S1).}
Normalized fiber length trajectories for each muscle across trials, colored by load. Patterns reflect physiological shortening/lengthening with robust estimation (narrow SD bands) across conditions.}
\label{fig:fiber_length_comparison}
\end{figure}
\subsection{Joint torque, muscle activation, and fiber length estimation across loading conditions}

To evaluate the consistency of model predictions across varying load conditions, we analyzed the joint torque, muscle activation, and fiber length trajectories using the PSO-calibrated model. Figures~\eqref{fig:torque_comparison}, \eqref{fig:activation_comparison}, and \eqref{fig:fiber_length_comparison} present time-normalized ensemble-averaged results for a representative subject (S01) across 0, 2, and 4~kg load conditions.
Model-predicted joint torque profiles showed close agreement with experimental data across all loading conditions. The predicted trajectories largely overlapped with the experimental mean $\pm$ SD bands, particularly during the flexion and extension phases. Slight deviations at torque peaks were observed, especially under higher loads.

Muscle activation profiles revealed a clear load-dependent modulation, with higher peak amplitudes and earlier onset times at increased loads. Activation timings for the biceps and triceps corresponded with expected movement phases across conditions.
Predicted muscle fiber lengths also varied with load, displaying consistent trends across trials. The trajectories exhibited smooth profiles with narrow inter-trial standard deviation bands, indicating stable estimation across repetitions with variable durations.
These observations highlight consistent modulation in neural and mechanical outputs in response to changing load, as reflected by the model’s output trajectories across all examined variables.
\subsection{Evaluation of calibration strategies}

To compare the performance of different MSK model calibration strategies, five approaches were evaluated using the PSO optimization framework: (i) load-specific calibration, (ii) fixed-global calibration, and three cross-load calibrations using (iii) fixed-0~kg, (iv) fixed-2~kg, and (v) fixed-4~kg data. 
Prediction accuracy was quantified using RMSE and Pearson correlation coefficient ($r$) between model-predicted and experimental joint torques across all subjects and load conditions.
For each strategy, the mean RMSE and $r$ were computed across all test conditions. To facilitate comparison, strategies were ranked based on these metrics, with lower RMSE and higher $r$ receiving better ranks. A combined score was calculated by summing the RMSE and correlation ranks.

Figure~\eqref{fig:calibration_strategy} presents the mean~$\pm$~SD of RMSE and $r$ values across all subjects for each calibration strategy. The top two strategies are visually highlighted, with yellow indicating the best and blue indicating the second-best performance. 
Load-specific calibration yielded the lowest RMSE and highest correlation values across all loads. Fixed-global calibration consistently ranked second, followed by the three fixed-load cross-calibration strategies. 
Paired \textit{t}-tests comparing each strategy to the load-specific reference condition revealed significant differences in RMSE and correlation for most cross-load strategies. Significance markers are indicated in the figure for reference.
\begin{figure*}[ht!]
  \centering
  \begin{subfigure}[t]{0.52\textwidth}
    \includegraphics[width=0.90\linewidth]{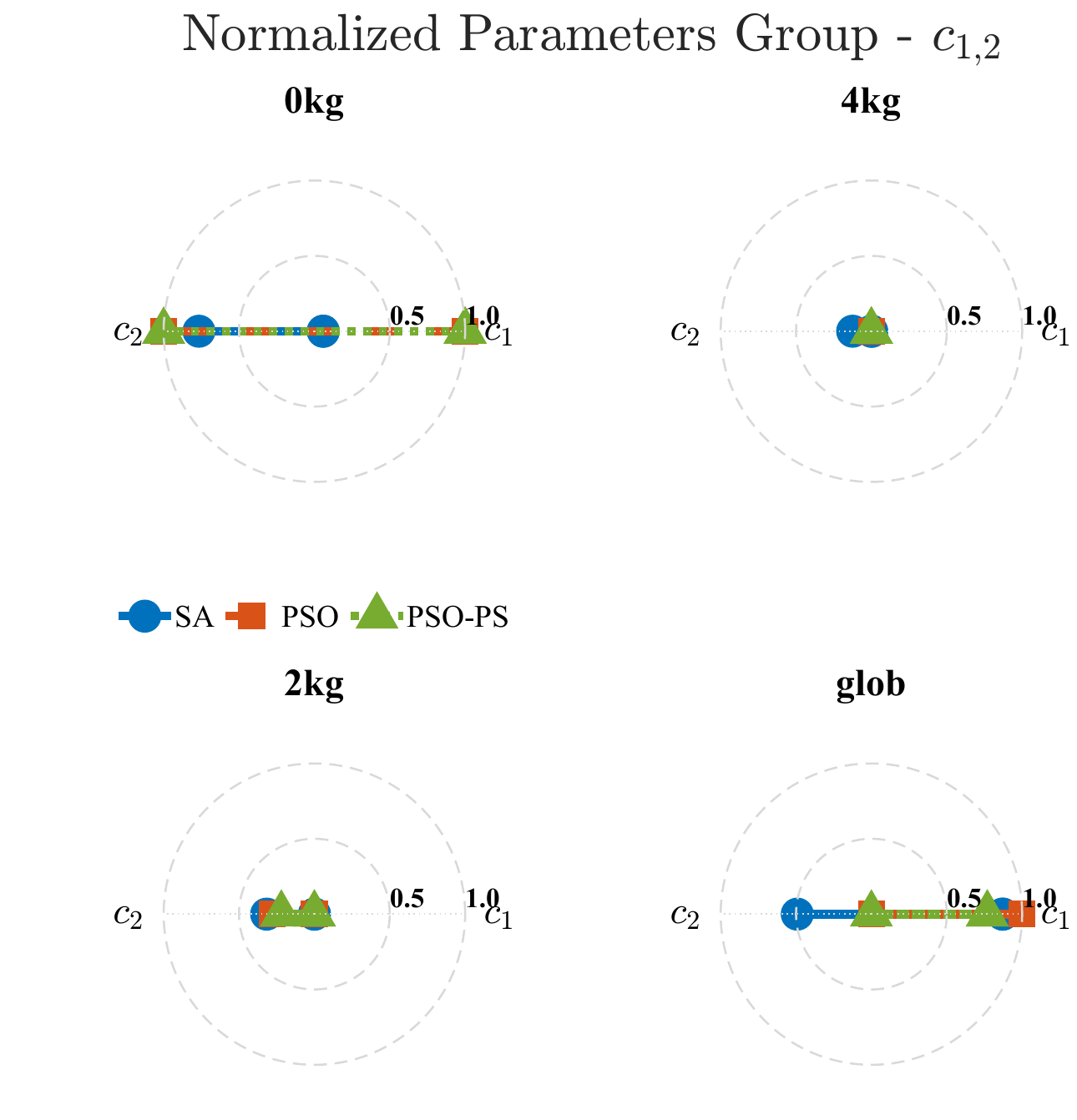}
    \caption{Group $\boldsymbol{c_{1,2}}$}
  \end{subfigure}
  \begin{subfigure}[t]{0.47\textwidth}
    \includegraphics[width=0.90\linewidth]{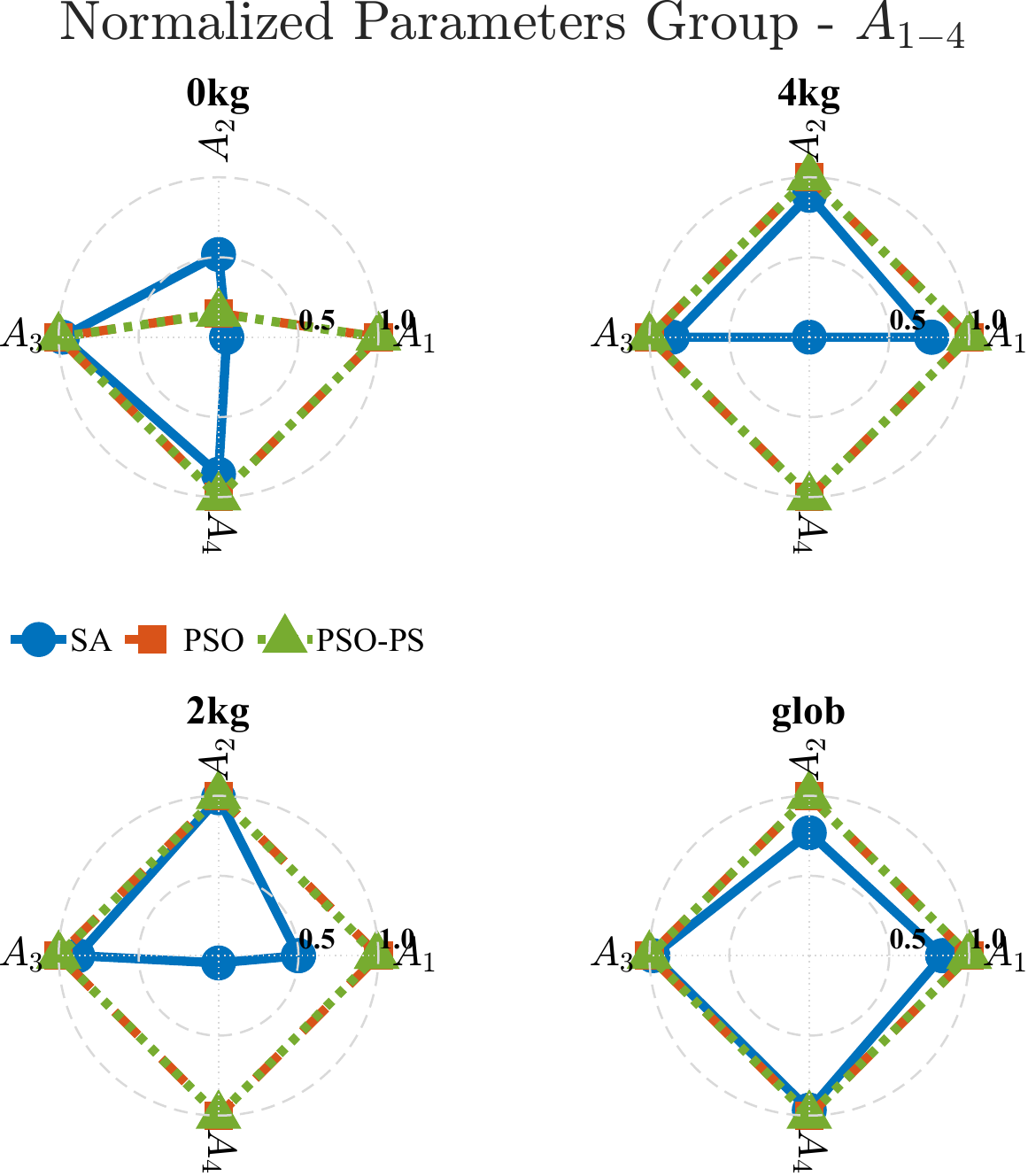}
    \caption{Group $\boldsymbol{A_{1-4}}$}
  \end{subfigure}

  \vspace{0.8em}
  
  \begin{subfigure}[t]{0.51\textwidth}
    \includegraphics[width=0.90\linewidth]{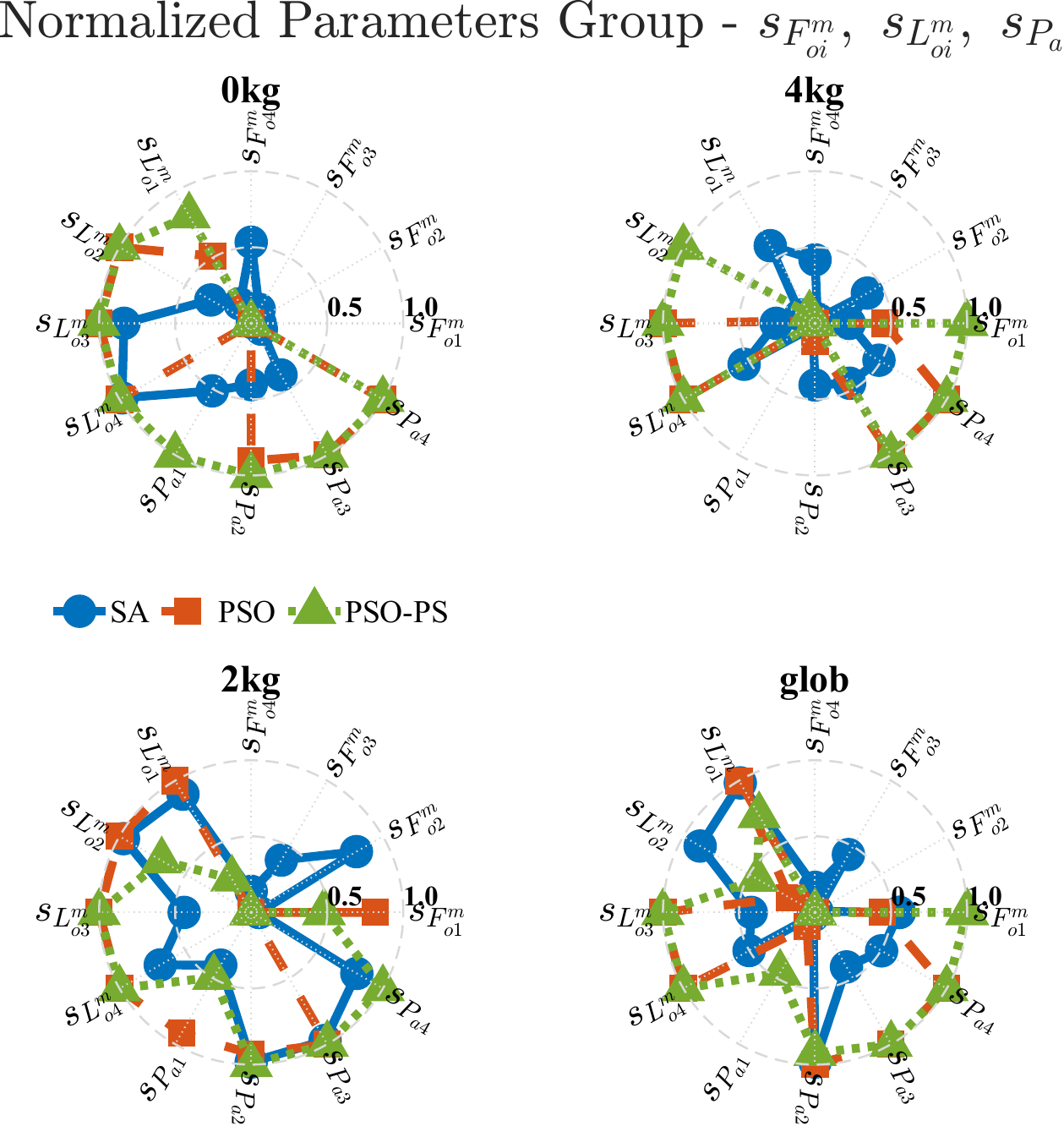}
    \caption{Group $\boldsymbol{s_{F^{m}_{oi}},\ s_{L^{m}_{oi}},\ s_{P_a}}$}
  \end{subfigure}
  \begin{subfigure}[t]{0.48\textwidth}
    \includegraphics[width=0.90\linewidth]{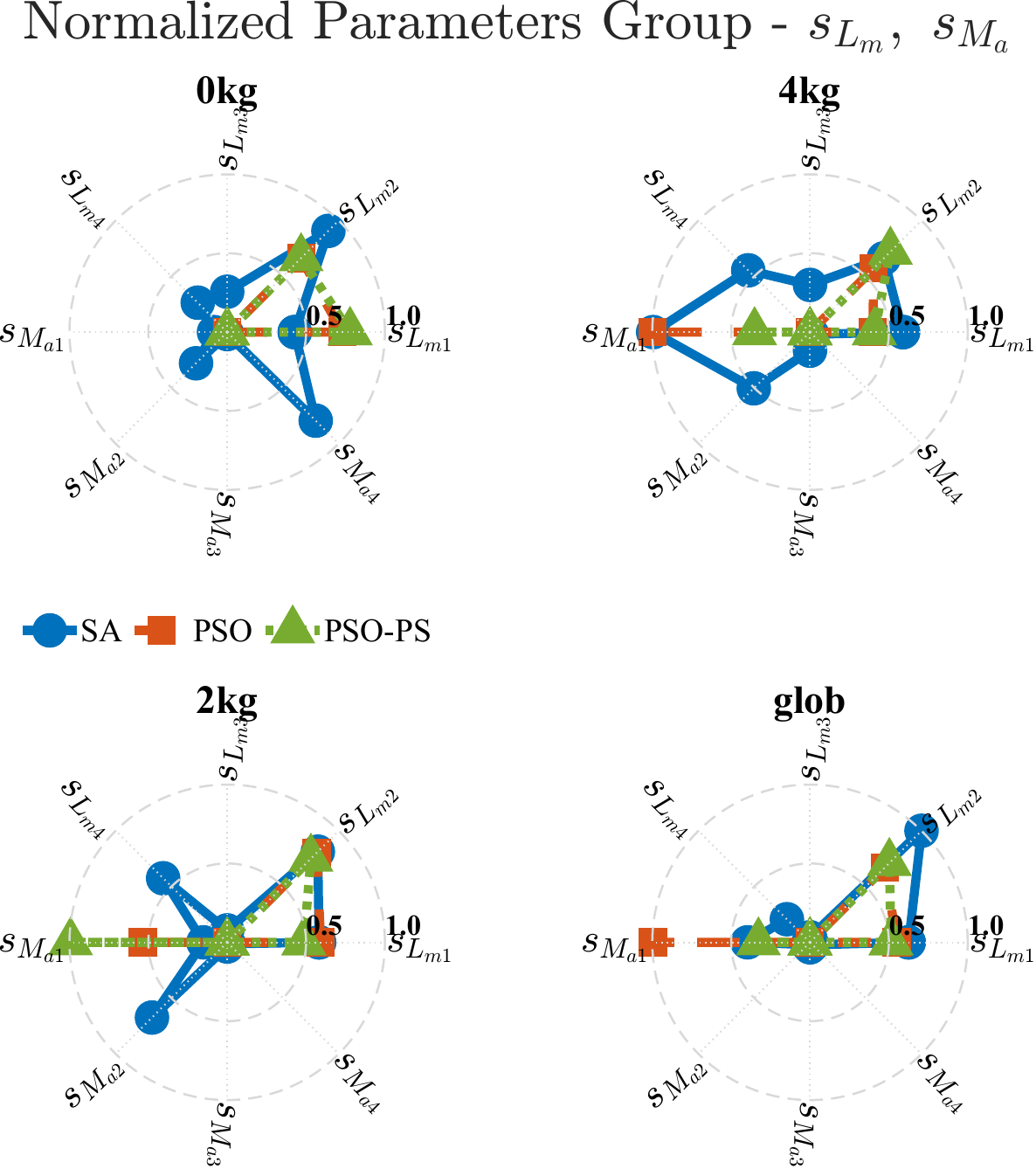}
    \caption{Group $\boldsymbol{s_{L_m},\ s_{M_a}}$}
  \end{subfigure}
  \caption{\textbf{Normalized parameter distributions across optimization methods and loads.}
Radar plots (a–d) show normalized values (0–1) of grouped musculoskeletal parameters for simulated annealing (SA), particle swarm optimization (PSO), and hybrid PSO-pattern search (PSO-PS) across all loading conditions (0, 2, 4~kg, and global). Each panel corresponds to a parameter group, illustrating calibration consistency and variability. Tighter clustering for PSO/PSO-PS, especially in scaling and geometry groups, highlights the robustness and physiological plausibility of population-based optimization.}
  \label{fig:radar_parameter_methods}
\end{figure*}
\begin{figure*}[ht]
\centerline{\includegraphics[width=0.95\linewidth]
{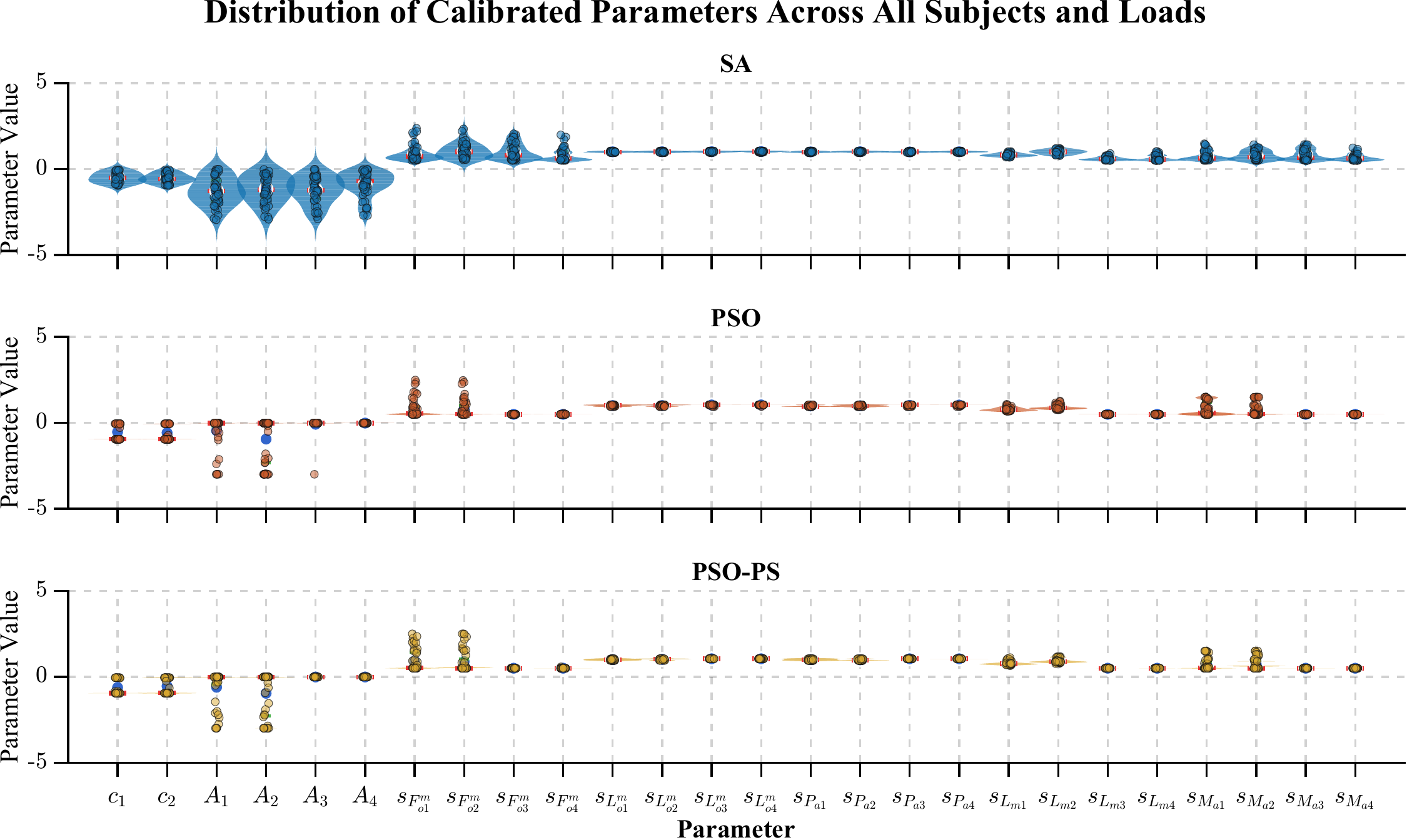}}
\caption{\textbf{Distribution of nominal values of calibrated MSK model parameters across all subjects and load conditions for each optimization method}: SA (top), PSO (middle), and PSO-PS hybrid (bottom). 
Each violin plot shows the full spread and density of parameter values; the red line denotes the median, the white dot is the mean, and green bars indicate the interquartile range. Compact violins reflect stable parameter estimation, while broader violins highlight parameters with higher variability across subjects and loads.}
\label{fig:violin_param_dist}
\end{figure*}
\subsection{Influence of optimization frameworks on parameter consistency}

To assess the effect of different optimization frameworks on the consistency of calibrated MSK parameters, normalized parameter distributions were visualized for a representative subject (S01) across all load conditions (0, 2, 4~kg, and global), grouped by physiological function (see Figure~\ref{fig:radar_parameter_methods}).
The PSO and PSO-PS frameworks produced tightly clustered parameter distributions across most parameter groups, including muscle strength scaling ($s_{F^{m}_{o}}$), optimal fiber length scaling ($s_{L^{m}_{o}}$), moment arm scaling ($s_{M_{a}}$), and the nonlinearity parameters ($A_{1}$--$A_{4}$). These distributions were relatively stable across all load conditions.

In contrast, the SA framework exhibited greater variability in several parameter groups, particularly under higher load conditions. For EMG-to-activation dynamics parameters ($c_1$, $c_2$), all three methods yielded similar distributions at lower loads. However, under global and high-load calibration, PSO and PSO-PS maintained more consistent parameter values than SA.
These trends reflect differences in parameter convergence behavior across optimization methods, as evident from the spread and overlap of the distributions across load conditions.
\subsection{Parameter Sensitivity and Stability Analysis}

To assess the influence of individual model parameters on joint torque estimation, a one-at-a-time (OAT) sensitivity analysis was conducted. Each parameter was perturbed by $+5\%$, and the resulting change in RMSE was computed relative to the baseline configuration. This process was repeated under three external load conditions: 0\,kg, 2\,kg, and 4\,kg.
Figure~\ref{fig:sensitivity_bar} illustrates the OAT sensitivity for a representative subject (S01), with parameters ranked by their RMSE impact. Among all parameters, $c_2$ and $s_{L_{m1}}$ showed the highest sensitivity, especially at 2\,kg and 4\,kg. Most others induced minimal RMSE changes, indicating localized sensitivity.
Figure~\ref{fig:sensitivity_heatmap} summarizes the average sensitivity results across all 11 subjects. The heatmaps display both nominal parameter values and the mean relative RMSE increase following perturbation. While sensitivity was low for most parameters (typically below 0.01), consistent elevation was observed for $c_1$, $c_2$, $s_{L_{m1}}$, and $s_{L_{m2}}$ across all loads. At 4\,kg, the maximum RMSE increase reached approximately 0.27. These findings highlight a small subset of load-sensitive parameters with disproportionate impact on torque prediction.

\vspace{-1em}
\subsection{Statistical Analysis of Calibrated Parameters}

\subsubsection{Distribution Across Loads and Methods}

To examine variability in calibrated parameters across subjects, loads, and optimization frameworks, violin plots were used to visualize the distributions of all 26 parameters (Figure~\ref{fig:violin_param_dist}). Separate panels represent results from SA, PSO, and PSO-PS.
Most parameters exhibited compact, unimodal distributions with closely aligned mean and median values, indicating stable convergence. However, certain parameters—particularly those related to muscle force scaling ($s_{F^{m}_{o2}}$, $s_{F^{m}_{o3}}$) and length scaling ($s_{L_{m2}}$, $s_{L_{m3}}$)—showed wider spread and occasional bimodality. In contrast, EMG-to-activation filter coefficients ($c_1$, $c_2$) and optimal fiber length scalings ($s_{L^{m}_{o}}$) demonstrated narrower and more consistent distributions. Among the methods, SA generally produced a broader parameter spread compared to PSO and PSO-PS.

\subsubsection{Cross-Method Sensitivity of Calibrated Parameters}

To evaluate the effect of the optimization algorithm on parameter estimates, a one-way ANOVA was performed for each parameter across all subjects and methods (SA, PSO, and PSO-PS) under each load condition (0\,kg, 2\,kg, 4\,kg, and global). The resulting $p$-values are presented in Figure~\ref{fig:anova_pvalues}, with statistically significant differences ($p < 0.05$) highlighted in orange.
Across all conditions, most parameters showed no significant differences between methods, reflecting general agreement across frameworks. However, a subset—including $A_2$, $A_3$, $s_{F^m_{o1}}$, $s_{F^m_{o12}}$, $s_{L^m_{o3}}$, $s_{P_{a2}}$, and $s_{L_{m2}}$—exhibited significant variability. Many of these correspond to muscle-strength or geometry parameters associated with antagonist muscles, such as the triceps long and lateral heads.
Notably, the global calibration condition exhibited the fewest parameters with significant differences across methods, suggesting improved consistency when parameter estimation is performed using aggregated load data.

\section{Discussion}\label{sec:discussion}

This study systematically investigates the impact of external mechanical load and optimization framework on EMG-driven musculoskeletal (MSK) model calibration for the elbow joint. A key contribution lies in demonstrating that load-aware calibration is essential for enabling MSK models to distinguish between kinematically similar movements that differ in mechanical demand—an important requirement in real-world and clinical applications.

Our findings reveal that several muscle-specific parameters, particularly maximum isometric force and optimal fiber length, vary systematically with increasing external load. This contradicts the commonly held assumption of load-invariant subject-specific parameters~\cite{colacino2012subject, falisse2016emg, crossley2025calibrated, wu2016subject}, and emphasizes the physiological necessity for load-specific tuning. Furthermore, we observed subtle modulations in EMG-to-activation dynamics and muscle geometry parameters, underscoring the adaptive nature of neuromuscular responses under changing load conditions.

Comparing three optimization frameworks—Simulated Annealing (SA), Particle Swarm Optimization (PSO), and hybrid PSO-pattern search (PSO-PS)—we found that population-based strategies (PSO, PSO-PS) delivered tighter, more physiologically plausible parameter estimates and better torque prediction, especially under higher loads. The hybrid PSO-PS consistently achieved the lowest calibration error and highest correlation, aligning with prior work advocating multi-stage global-local strategies for high-dimensional optimization~\cite{jiang2025personalized, schutte2005evaluation}.

Sensitivity analysis revealed that a small subset of parameters—primarily those governing force and length scaling and activation dynamics—had the greatest influence on torque prediction. This is consistent with previous findings~\cite{reed2015optimising, hinson2022sensitivity, hosseini2022uncertainty} and suggests that future calibration efforts could focus on this core parameter subset to reduce computational burden.
Among calibration strategies, load-specific calibration yielded the best prediction accuracy, followed by fixed-global calibration. Cross-load strategies performed worst when load mismatch occurred. These results reinforce that \textbf{adaptive, load-aware calibration is critical} for achieving robust generalization across task contexts.

While our framework demonstrates strong performance in quasi-static elbow flexion-extension tasks using surface EMG from four muscles, generalization to dynamic, multi-joint movements or additional musculature remains to be explored. Moreover, although the hybrid PSO-PS method improves accuracy, its computational cost may pose challenges for real-time applications. Future research should extend load-aware calibration to dynamic and functionally diverse tasks, incorporate probabilistic or learning-based inference for efficiency, and consider sensor fusion (e.g., EMG + IMU) for broader applicability.

In summary, our results establish that load-specific, hybrid-optimized calibration enhances both predictive performance and physiological interpretability in EMG-driven MSK models. This load-aware framework equips biomechanical models with the capability to disambiguate visually identical yet mechanically distinct movements, addressing a key challenge for their deployment in naturalistic, occupational, and assistive settings. 
\section{Conclusion}
\label{sec:conclusion}

This study demonstrates the necessity and effectiveness of load-aware calibration in EMG-driven musculoskeletal (MSK) models of the elbow. By calibrating model parameters under varying external loads, we show that load-specific tuning enables accurate joint torque estimation across mechanically distinct yet visually similar tasks—an essential capability for real-world applications.
Our analysis reveals that key parameters, particularly those related to muscle strength and geometry, adapt systematically to load, challenging the assumption of load-invariant calibration. Optimization frameworks such as PSO and hybrid PSO-PS provided consistent, physiologically plausible parameter estimates across subjects and conditions. Sensitivity analysis further identified a subset of influential parameters that can be prioritized for efficient calibration.
These findings support the broader use of adaptive MSK models in applications such as assistive devices and rehabilitation, where robustness across task contexts is critical. Future work will extend this approach to dynamic, multi-joint tasks and explore scalable strategies for real-time deployment.

\section*{Acknowledgment} \label{Acknowledgment}
The authors would like to thank Anant Jain from Indian Institute of Technology (IIT) Delhi for helping in data collection. This study is partly supported by the Joint Advanced Technology Centre (JATC) (Grant no.: RP03830G) and the I-Hub Foundation for cobotics (IHFC) section-8 company (Grant no.: GP/2021/RR/010) at IIT Delhi, sponsored by the Ministry of Education (MoE), Govt. of India.

\appendix
\onecolumn
\section*{Appendix A \\ OAT Parameter Sensitivity}\label{OAT_S01_results}
\begin{figure*}[ht!]
    \centering
    \includegraphics[width=0.60\textwidth]{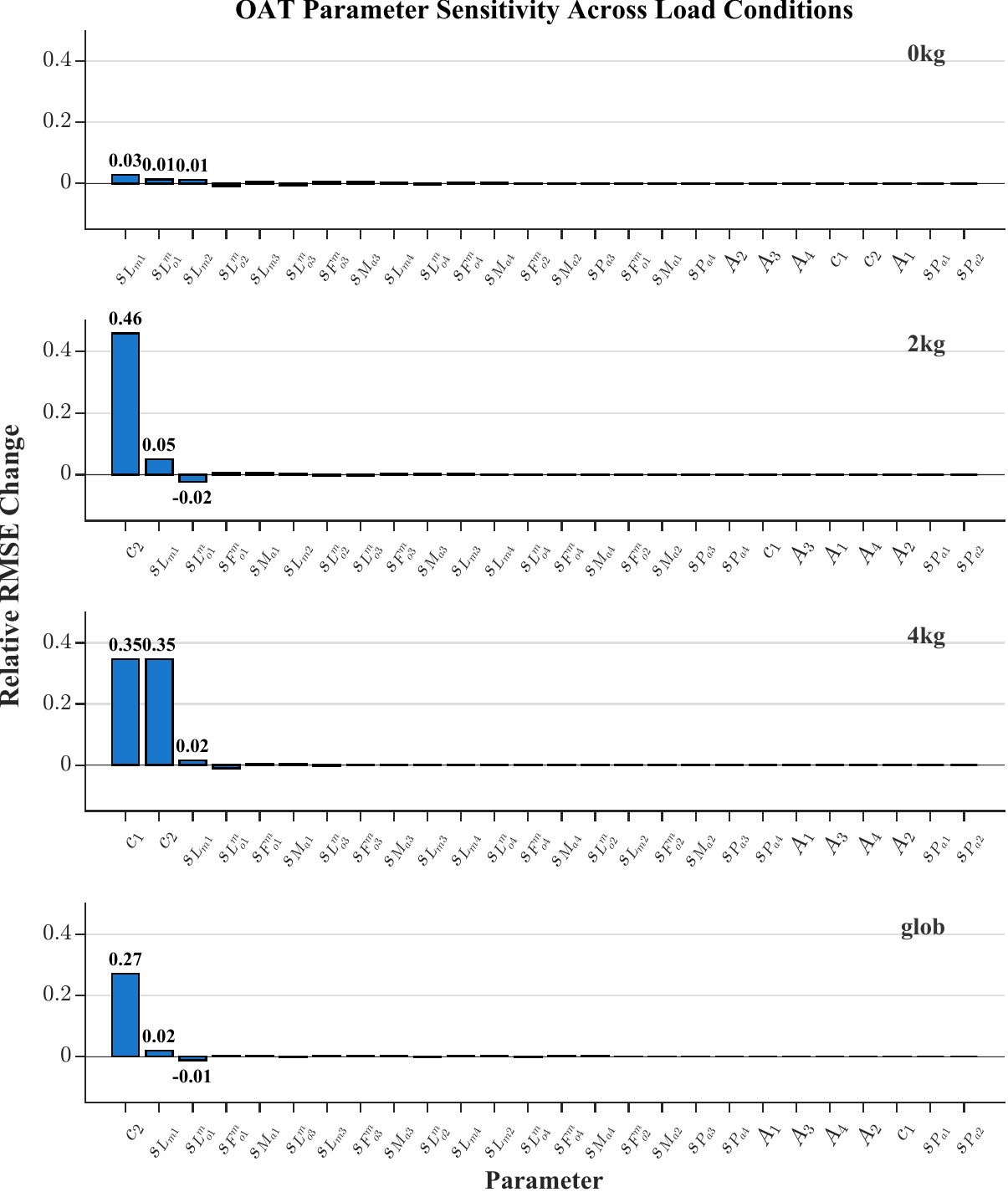}
    \caption{\textbf{OAT Parameter Sensitivity Across Load Conditions.} Each panel represents the impact of individual musculoskeletal model parameters on joint torque prediction accuracy for a representative subject, S01, under four external load conditions (0\,kg, 2\,kg, 4\,kg, and global). Each bar indicates the relative increase in RMSE caused by a $+5\%$ perturbation of the corresponding parameter.}
    \label{fig:sensitivity_bar}
\end{figure*}
\vspace{-3em}
\begin{figure*}[ht!]
    \centering
    \includegraphics[width=0.70\textwidth]{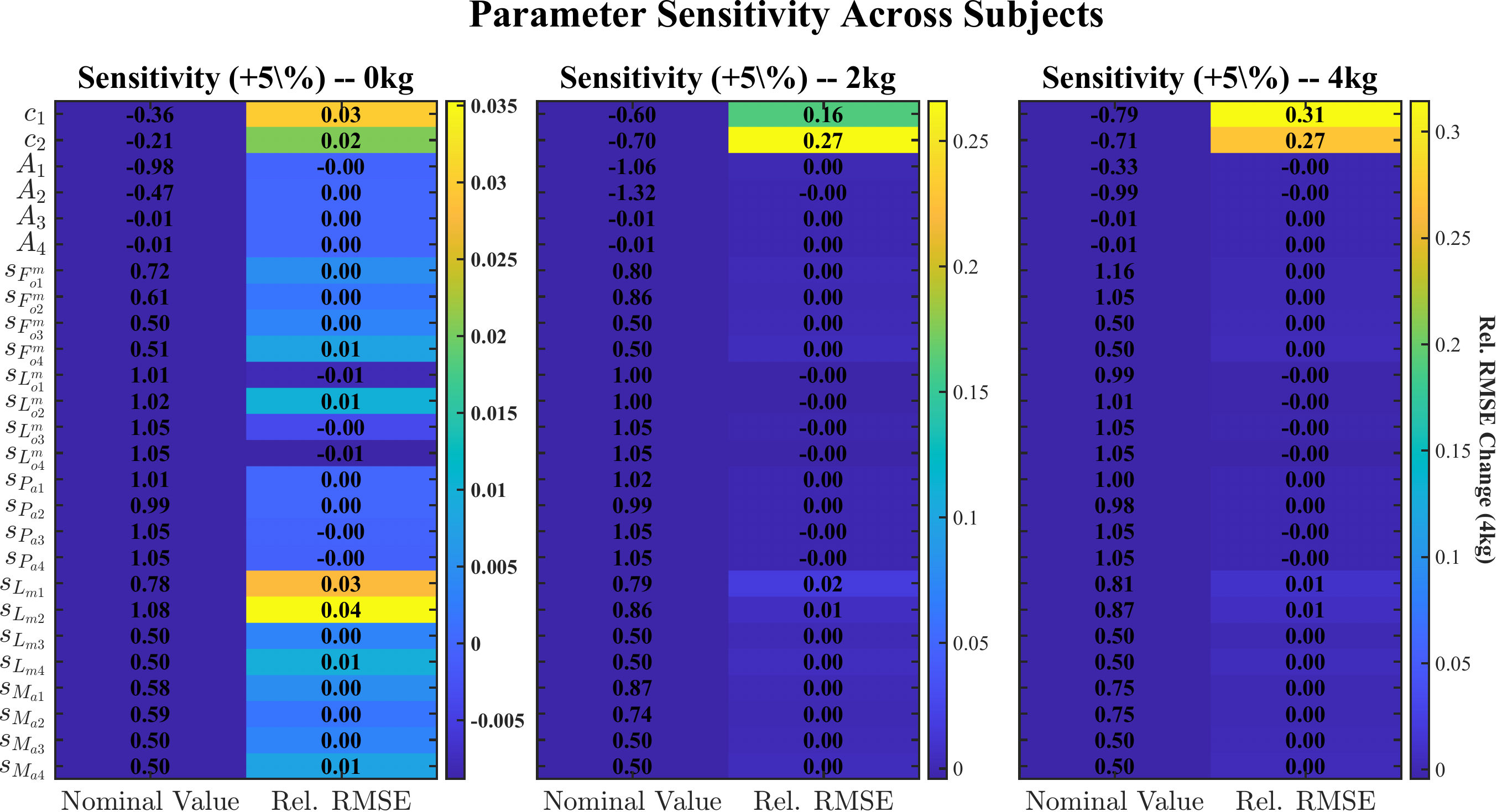}
    \caption{\textbf{Aggregated Parameter Sensitivity Across Subjects and Load Conditions.} Relative RMSE changes following $+5\%$ perturbation of each parameter are shown across all load conditions. Parameters such as $c_2$ and $s_{L_{m1}}$ show consistently high sensitivity, corroborating the trends observed in Figure~\ref{fig:sensitivity_bar}.}
    \label{fig:sensitivity_heatmap}
\end{figure*}
\clearpage
\section*{Appendix B \\Cross-Method ANOVA $p$-Values}\label{fig_anova_results}

\begin{figure*}[ht]
    \centering
    \includegraphics[trim=0cm 2cm 0cm 2cm, clip=true,width=0.70\textwidth]{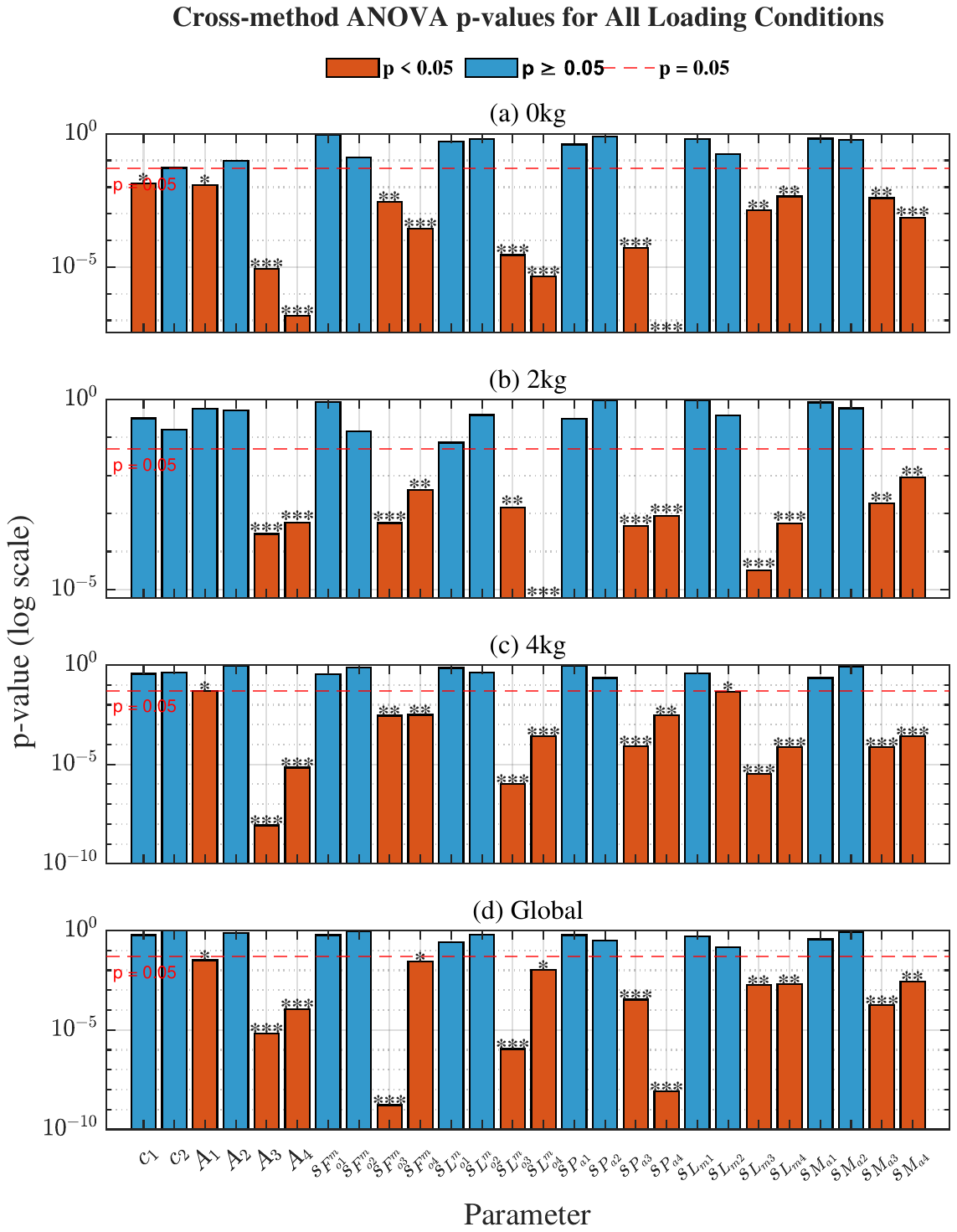}
    \caption{\textbf{Cross-Method ANOVA $p$-Values.} $p$-values for ANOVA tests comparing calibrated parameters across optimization methods (SA, PSO, PSO-PS) for each load condition. Horizontal dashed line indicates significance threshold ($p = 0.05$, shown in log scale). Significant differences are marked: * $p<0.05$, ** $p<0.01$, *** $p<0.001$.}
    \label{fig:anova_pvalues}
\end{figure*}
\twocolumn 
\section*{Appendix C \\Pseudocode for Optimization of MSK Parameters}\label{optim_pseudocode}

This appendix presents the complete pseudocode used for musculoskeletal (MSK) parameter calibration. The main loop iterates over optimization methods and load conditions and internally calls one of three optimization algorithms depending on the selected method.

\noindent
\subsection{Main Optimization Loop for EMG-Driven MSK Parameter Calibration}\label{app:main_pseudocode}

\begin{table}[ht!]
\renewcommand{\arraystretch}{1.1}
\begin{tabular}{p{0.96\linewidth}}
\toprule
\textbf{Input:} Data $\mathcal{D}_{\text{all}}$, parameter bounds $[\text{LB}, \text{UB}]$, optimization methods $\mathcal{M}$, load conditions $\mathcal{L}$ \\
\textbf{For each} method $m \in \mathcal{M}$ \textbf{do} \\
\quad \textbf{For each} load condition $\ell \in \mathcal{L}$ \textbf{do} \\
\quad\quad Extract data $\mathcal{D}_\ell$ \\
\quad\quad Define objective function $\mathcal{J}_\ell(\mathbf{x}) \leftarrow$ joint torque error \\
\quad\quad Solve: \hspace{0.5em} $\mathbf{x}_\ell^* = \arg \min\limits_{\mathbf{x} \in [\text{LB}, \text{UB}]} \mathcal{J}_\ell(\mathbf{x})$ using method $m$ (see Section~B) \\
\quad\quad Store optimal parameters $\mathbf{x}_\ell^*$ and cost $\mathcal{J}_\ell(\mathbf{x}_\ell^*)$ \\
\quad \textbf{End For} \\
\quad Save $\{ \mathbf{x}_\ell^* \}_{\ell \in \mathcal{L}}$ for method $m$ \\
\textbf{End For} \\
\bottomrule
\end{tabular}
\end{table}
\noindent
\subsection{Optimization Methods Used in the Main Loop}\label{app:optimizer_pseudocode}

\noindent The following routines implement the internal optimizer in Line 6 of the main loop pseudocode. The selected method $m \in \{\texttt{SA}, \texttt{PSO}, \texttt{PSO-PS}\}$ determines which algorithm is executed.

\begin{table}[ht!]
\renewcommand{\arraystretch}{1.1}
\caption*{\textbf{B.1 Simulated Annealing (SA)}}
\begin{tabular}{p{0.96\linewidth}}
\toprule
\textbf{Input:} Objective function $\mathcal{J}(\mathbf{x})$, bounds $[\text{LB}, \text{UB}]$ \\
\textbf{Initialize:} Random guess $\mathbf{x}_0 \sim \mathcal{U}(\text{LB}, \text{UB})$ \\
\textbf{Set:} Simulated annealing options (e.g., max iterations, objective limit) \\
\textbf{Run:} $\mathbf{x}^* \gets \texttt{simulannealbnd}(\mathcal{J}, \mathbf{x}_0, \text{LB}, \text{UB})$ \\
\textbf{Output:} Optimized parameters $\mathbf{x}^*$ and cost $\mathcal{J}(\mathbf{x}^*)$ \\
\bottomrule
\end{tabular}
\end{table}
\vspace{-1em}
\begin{table}[ht!]
\renewcommand{\arraystretch}{1.1}
\caption*{\textbf{B.2 Particle Swarm Optimization (PSO)}}
\begin{tabular}{p{0.96\linewidth}}
\toprule
\textbf{Input:} Objective function $\mathcal{J}(\mathbf{x})$, bounds $[\text{LB}, \text{UB}]$ \\
\textbf{Set:} Swarm size, max iterations, search space \\
\textbf{Run:} $\mathbf{x}^* \gets \texttt{particleswarm}(\mathcal{J}, \dim(\mathbf{x}), \text{LB}, \text{UB})$ \\
\textbf{Output:} Optimized parameters $\mathbf{x}^*$ and cost $\mathcal{J}(\mathbf{x}^*)$ \\
\bottomrule
\end{tabular}
\end{table}
\vspace{-1em}
\begin{table}[ht!]
\renewcommand{\arraystretch}{1.1}
\caption*{\textbf{B.3 Hybrid PSO-Pattern Search (PSO-PS)}}
\begin{tabular}{p{0.96\linewidth}}
\toprule
\textbf{Input:} Objective function $\mathcal{J}(\mathbf{x})$, bounds $[\text{LB}, \text{UB}]$ \\
\textbf{Step 1:} Run PSO to get coarse solution: \\
\quad $\mathbf{x}_{\text{PSO}} \gets \texttt{particleswarm}(\mathcal{J}, \dim(\mathbf{x}), \text{LB}, \text{UB})$ \\
\textbf{Step 2:} Refine with pattern search: \\
\quad $\mathbf{x}^* \gets \texttt{patternsearch}(\mathcal{J}, \mathbf{x}_{\text{PSO}}, \text{LB}, \text{UB})$ \\
\textbf{Output:} Final optimized parameters $\mathbf{x}^*$ and cost $\mathcal{J}(\mathbf{x}^*)$ \\
\bottomrule
\end{tabular}
\end{table}
\section*{Appendix D \\Detailed musculoskeletal model expressions}\label{MSK_expressions_detailed}

\subsection{Activation Module Parameter Constraints}\label{app:activation-constraints}
\begin{equation}
\begin{aligned}
& |c_1| < 1,\quad |c_2| < 1 \\
& \beta_1 = c_1 + c_2,\quad \beta_2 = c_1 c_2 \\
& \alpha - \beta_1 - \beta_2 = 1
\end{aligned}
\end{equation}

\subsection{Musculotendon Kinematics: Expanded Polynomial Forms}\label{app:kinematics-poly}

\begin{equation}\label{eq:muscle_length_regression}
    \textit{L}_{mj}[n] = s_{L_{mj}} \left( \kappa_j + \sum_{i=1}^{2} \left( y_n q_i^n + y_{n-1} q_i^{n-1} + \cdots + y_1 q_i \right) \right)
\end{equation}

\begin{equation}\label{eq:moment_arm_regression}
    \textit{M}_{aj}[n] = s_{L_{mj}} \left( r_j + x_n q_i^n + x_{n-1} q_i^{n-1} + \cdots + x_1 q_i \right)
\end{equation}

where $\textit{L}_{mj}$ denotes the length of the $j^{\text{th}}$ muscle in millimeters, and $\textit{M}_{aj}$ denotes the moment arm of the $j^{\text{th}}$ muscle at a specific degree of freedom. The variable $q_i$ represents the joint angle in degrees, while $\kappa_j$ and $r_j$ is the constant portion of the $j^{\text{th}}$ muscle length and moment arm respectively. The coefficients $y_i$ and $x_i$ correspond to the polynomial terms for muscle length and moment arm estimation, respectively, with $y_i$ typically associated with elbow movements (e.g., flexion/extension), and $x_i$ with other upper limb motions \cite{pigeon1996moment}.

\subsection{Muscle contraction dynamics expressions in detail}\label{app:contraction-details}
\textbf{Active Force-Length:}
\begin{equation}
\bar{f}_L^{a}(\bar{l}_j^{m}[n]) = \exp\left(-\frac{(\bar{l}_j^{m}[n]-1)^2}{\gamma}\right)
\end{equation}

\textbf{Normalized Force-Velocity:}
\begin{equation}
    \bar{f}_V(\bar{v}_j^{m}[n])=
    \left\{\begin{aligned}
    &\frac{(v+1)}{1+{A_f^{-1}}}; \quad v<-1 \\
    &\frac{(v+1)}{1-{v}{A_f^{-1}}}; \quad-1 \leq v<0 \\
    &\frac{\left(2+{2}{A_f^{-1}}\right) v \bar{F}_{\text {len }}^M+\bar{F}_{\text {len }}^M-1}{\left(2+{2}{A_f^{-1}}\right) v+\bar{F}_{\text {len }}^M-1}; \quad 0 \leq v < \psi_{2} \\
    &\psi_{0} \left(\frac{\left(1+\frac{1}{A_f}\right) \bar{F}_{\text {len }}^M v}{10\left(\bar{F}_{\text {len }}^M-1\right)}+ \psi_{1} \right), \quad v > \psi_{2} 
    \end{aligned}\right.
\end{equation}
where $\psi_{0}$, $\psi_{1}$, and $\psi_{2}$ are given as, 
\begin{equation}
    \begin{aligned}
        &\psi_{0} = \frac{\bar{F}_{\text {len }}^M}{20\left(\bar{F}_{\text {len }}^M-1\right)}\\
        &\psi_{1} = 18.05 \bar{F}_{\text {len }}^M-18\\
        &\psi_{2} = \frac{10\left(\bar{F}_{\text {len }}^M-1\right)\left(0.95 \bar{F}_{\text {len }}^M-1\right)}{\left(1+\frac{1}{A_f}\right) \bar{F}_{\text {len }}^M}
    \end{aligned}
\end{equation}

\textbf{Passive Force-Length:}
\begin{equation}
    \bar{f}_L^{p}(\bar{l}_j^{m}[n]) = 
    \left\{\begin{aligned}
        &1+\frac{k^{\text{PE}}}{\varepsilon_0^M}\left(\bar{l}^M-\left(1+\varepsilon_0^M\right)\right); \bar{l}^M>1+\varepsilon_0^M \\
        &\frac{\exp\left({k^{\text{PE}}\left(\bar{l}^M-1\right) \varepsilon_0^K}\right)}{\exp\left({k^{R_E}}\right)}; \;\; \bar{l}^M \leq 1+\varepsilon_0^M
    \end{aligned}\right.
\end{equation}

\clearpage
\section*{References}

\begin{IEEEbiography}[{\includegraphics[width=1in,height=1.25in,clip,keepaspectratio]{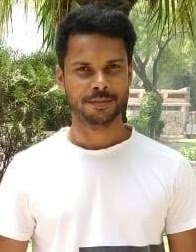}}]{Rajnish Kumar} received the B.E. degree from M. S. Ramaiah Institute of Technology, Bangalore, India, in 2014, and the M.Tech. degree from the Indian Institute of Technology (IIT) Delhi, India, in 2020. He is currently pursuing the Ph.D. degree with the Department of Applied Mechanics at IIT Delhi. His research focuses on biomechanics, neuro-musculoskeletal modeling, and the simulation of human motion, with applications in physics-informed machine learning, wearable soft robotic assistive devices, and neurorehabilitation. More information is available at: \href{https://www.linkedin.com/in/rajnishkuram}{www.linkedin.com/in/rajnishkuram}.
\end{IEEEbiography}
\begin{IEEEbiography}[{\includegraphics[width=1in,height=1.25in,clip,keepaspectratio]{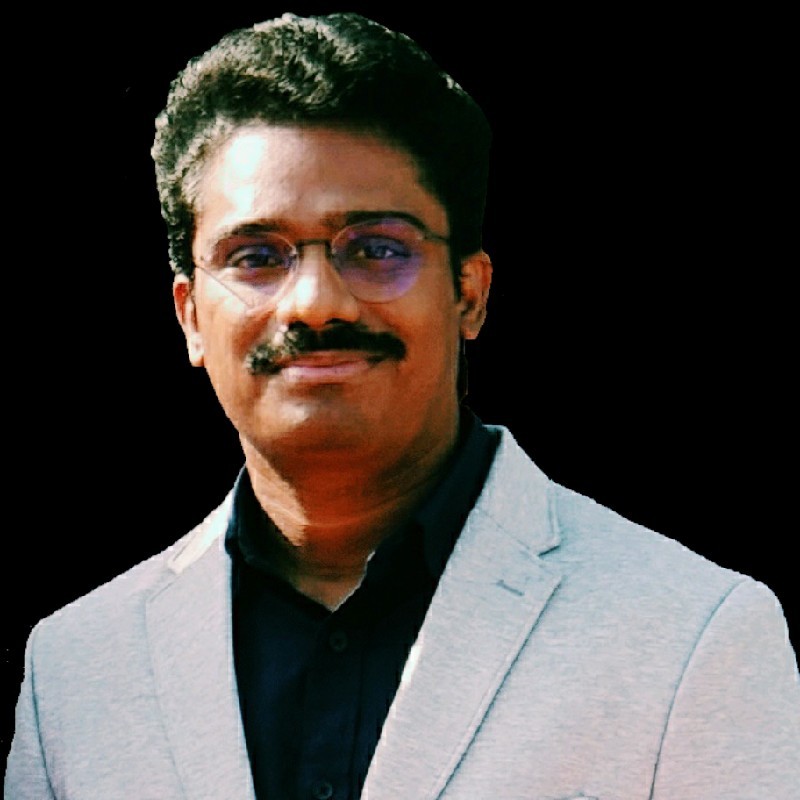}}]{Suriya Prakash Muthukrishnan} is an Associate Professor in the Department of Physiology at the All India Institute of Medical Sciences (AIIMS), New Delhi, India. His research interests include: (1) elucidating the neurophysiological mechanisms underlying human cognition, (2) identifying cost-effective biomarkers for diagnosis and prognosis of brain disorders, (3) developing novel treatment strategies to modulate neural networks in brain disorders, and (4) designing wearable devices for muscle power augmentation in both healthy individuals and patients with paralysis or paresis.
\end{IEEEbiography}
\begin{IEEEbiography}[{\includegraphics[width=1in,height=1.25in,clip,keepaspectratio]{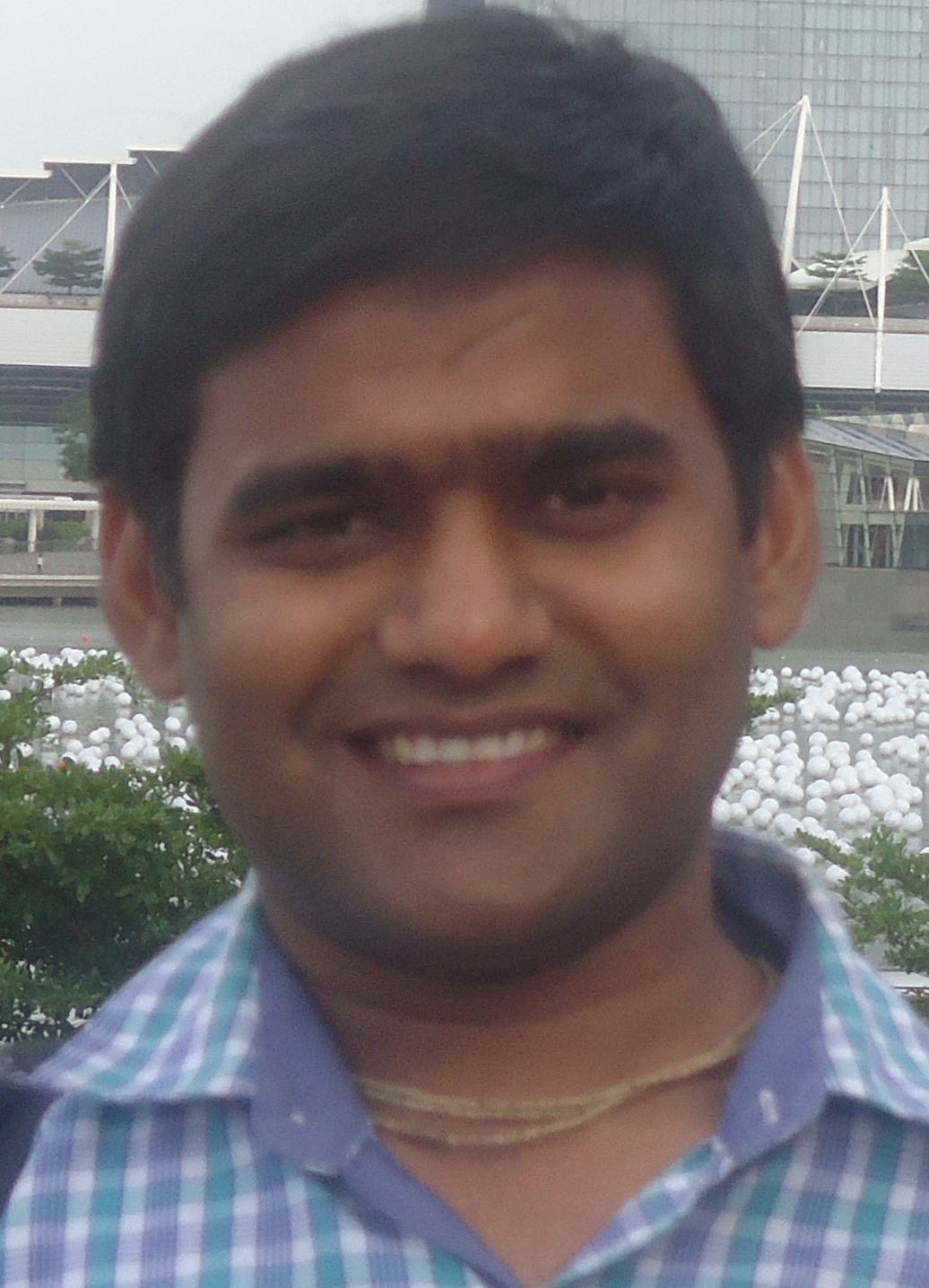}}]{Lalan Kumar} received his B.Tech. degree in Electronics Engineering from the Indian Institute of Technology (BHU) Varanasi, India, in 2008, and his Ph.D. degree from the Indian Institute of Technology Kanpur, India, in 2015. He is currently an Associate Professor in the Department of Electrical Engineering at the Indian Institute of Technology Delhi, New Delhi, India. Prior to this, he served as an Assistant Professor at IIT Bhubaneswar, India, and as a Research Fellow at Nanyang Technological University, Singapore. He also worked as a Software Engineer in the Multimedia Team at Motorola, Bengaluru, India, from 2008 to 2009. His research interests include brain source localization, brain–computer interfaces, and air-writing recognition. Detailed information is available at:\href{http://web.iitd.ac.in/~lalank/}{web.iitd.ac.in/~lalank/}.
\end{IEEEbiography}
\begin{IEEEbiography}[{\includegraphics[width=1in,height=1.25in,clip,keepaspectratio]{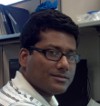}}]{Sitikantha Roy} received the B.E. degree from the Indian Institute of Engineering Science and Technology, Kolkata, India, in 2001, the M.Sc. degree from the Indian Institute of Science, Bengaluru, India, in 2004, and the Ph.D. degree from Utah State University, Logan, UT, USA, in 2007. He is currently Professor in the Department of Applied Mechanics at the Indian Institute of Technology Delhi, New Delhi, India. He has previously worked as a Postdoctoral Fellow at Johns Hopkins University, Baltimore, MD, USA, and the University of Colorado, Boulder, CO, USA. His research interests span brain and neuro-musculoskeletal biomechanics, human motion analysis, multi-flexible body dynamics, soft robotics, and data-/physics-driven AI/ML methods. More information is available at:\href{https://sites.google.com/view/sitilab-iitd/}{BSR Lab Website}.
\end{IEEEbiography}

\end{document}